\documentclass[twocolumn,showpacs,preprintnumbers,amsmath,amssymb]{revtex4}
\usepackage{graphicx}
\usepackage{dcolumn}
\usepackage{bm}
\newcommand{\sech}{\mbox{sech}}
\begin{document} 
\title{\bf Thermal diffusion of supersonic solitons
in an anharmonic chain of atoms}

\author{Edward Ar\'evalo and Franz G. Mertens}
\affiliation{Physikalisches Institut, Universit\"at Bayreuth,
D-95440 Bayreuth, Germany}
\author{ Yuri Gaididei}
\affiliation{Institute for Theoretical Physics, 252143 Kiev,
Ukranie}
\author{A. R. Bishop}
\affiliation{Theoretical Division and Center for Nonlinear
Studies, Los Alamos National Laboratory, Los Alamos, New Mexico 87545}
\date{\today}
\begin{abstract}
We study the non-equilibrium diffusion dynamics of supersonic lattice
solitons in a classical chain of 
atoms with nearest-neighbor interactions coupled to a heat bath.
As a specific example we choose an interaction with cubic
anharmonicity.  The coupling between the system and a thermal bath with
a given temperature is made by adding noise, delta-correlated
in time and space, and damping to the
set of discrete equations of motion. Working in the continuum limit
and changing to the sound velocity frame we derive a Korteweg-de
Vries equation with noise and damping.
We apply a collective coordinate approach which yields two stochastic
ODEs which are solved approximately by a perturbation analysis. This
finally yields analytical expressions for the
variances of the soliton position and velocity. We perform 
Langevin dynamics simulations for the original discrete system which
confirm the predictions of our
analytical calculations, namely noise-induced superdiffusive behavior
which scales with the temperature and depends strongly on the initial
soliton velocity. A normal diffusion behavior is observed for
solitons with very low energy where  the noise-induced
phonons also make a significant contribution to the soliton diffusion.
\end{abstract} 
\maketitle
\section{Introduction}

Nonlinear one-dimensional lattice dynamics, namely propagation of coherent
excitations in monatomic chains modeling discrete microscopic structures,
is associated with several important problems in physics. Among these
excitations are solitary waves, which for simplicity are called
here solitons. These solitons can be
supported by chains with realistic interaction
potentials between the particles \cite{Lomdahl,Lawrence}. They are supersonic non-topological 
collective excitations. In spite of their relative simplicity, the solitons clarify many
features of molecular chains \cite{Toda,Pnevmatikos,Davydov,Collins,Hochstrasser1}. 
For example,  due to their  robust character, lattice solitons
have been used to model the energy transport in polypeptide chains
in muscle proteins \cite{Yomosa,Perez,Perez1} or the energy transport
in DNA \cite{Muto}. Numerical simulations at realistic
temperatures for transport in proteins have shown that lattice
solitons can propagate over long distances in a chain with the Lennard-Jones potential
\cite{Perez}. Moreover, the lattice solitons are more stable than
Davydov solitons if collisions between the two types of solitons are
considered \cite{Perez1}. There is no clear evidence that lattice
solitons like a Toda type, which are non-topological, 
can exist in thermal equilibrium. This holds  both for static
properties, like the specific heat, and for dynamics quantities, like
the dynamic form factor (Fourier Transform of the displacement
autocorrelation) \cite{Neuper}. On the other hand, there exists evidence from real
experiments that strain solitons can be generated and
observed in non-linear elastic rods \cite{samsonov}. These solitons in
some cases can be described by  Korteweg-de Vries 
(KdV) type solitons, which are non-topological.

To our knowledge there are no previous
analytical studies supported by Langevin simulations about
non-topological lattice soliton diffusion in anharmonic 
monatomic chains of particles with nearest-neighbor
interactions. There are many studies on  stochastic partial differential
equations, in particular stochastic KdV-type equations have been
extensively studied numerically and analytically
\cite{KonotopVazquez,Wadati,Wadati2,Herman,Iizuka,ScalerandiRomano}
due to the integrability of the KdV equation. In fact,
the KdV equation is a good approximation to describe analytically the
dynamics of lattice solitons  on a monatomic chain with
nearest-neighbor interaction and cubic anharmonicity if the soliton
velocity is very close to the sound velocity (very-low-energy
solitons) \cite{Pnevmatikos,Remoissenet}. Notice that for a polynomial
potential, namely harmonic term plus cubic or/and quartic anharmonicity, 
the one-soliton solution of the KdV equation is known analytically, while for more
realistic interaction potentials like  Lennard-Jones or Morse there
are no exact soliton solutions. In the more general context of lattice
systems, there are a few analytical studies about diffusion of coherent
lattice excitations, viz. stochastic vortex dynamics in
two-dimensional easy-plane ferromagnets \cite{kampetter} or soliton
diffusion on the classical, isotropic Heisenberg chain
\cite{mathias1,mathias}. 

The aim of this work is to provide an approximate
analytical description of the soliton diffusion dynamics in a
monatomic chain with a cubic anharmonicity
under thermal fluctuations. For this purpose we generate a single soliton
which has an energy far greater than $k_B\,T$, where $k_B$ is the 
Boltzmann constant, and $T$ is the temperature of a thermal bath.
This soliton propagating on a chain in contact with the thermal bath
shows a diffusive behavior. 
We consider this soliton diffusion dynamics during the thermalization
process of the system up to times when the system energy  
has relaxed nearly to its stationary value. This means that we
study the non-equilibrium diffusion dynamics of lattice
solitons on  anharmonic chains subject to thermal fluctuations.

In order to perform the coupling between the system and a
thermal bath with a given temperature 
an additive noise term, providing energy input, is added to the discrete
equations of motion. This term has to be
balanced by a damping term providing energy dissipation. Here, we
suggest as a damping term  the so-called  hydrodynamical 
damping \cite{Landau} which is extensively used in, e.g., elasticity
theory. Notice that this type of damping is due to irreversible
processes taking place within the system.
The corresponding noise term, which fulfills the fluctuation-dissipation
theorem, takes the form of a discrete gradient of Gaussian white noise
delta-correlated in space and time. A similar Langevin-type equations
has previously been considered in the context of mesoscopic Langevin
dynamics \cite{Schmittmann}.

We notice that our system in the continuum limit 
can be approximated by a noisy KdV-Burgers-type equation 
\cite{nozaki,Orlowski}. So in this case we can use the one-soliton
solution of the KdV equation not only as initial condition of our
discrete system but also in our analytical approach in the continuum
limit. Notice that the shape of broad KdV solitons
tends to be identical to the shape of broad supersonic lattice
solitons \cite{Pnevmatikos}. In this work we apply a generalized
traveling  wave ansatz combined with a collective coordinate formalism
in the framework of the KdV equation  as an analytical approach 
to study the  diffusion of lattice solitons.

In the next section we present the equations of motion of our discrete
system. From this we formulate a set of  stochastic equations of motion
by adding noise and damping. Next, we apply the
continuum limit and derive a form of noisy KdV-Burgers equation. 
In section \ref{seccollec} we apply a collective coordinate approach which yields
analytical expressions for the thermal averages and variances of the soliton 
position and velocity. In section \ref{simulations}, we compare our analytical
predictions with the results from Langevin dynamics simulations for the
original discrete system. Our conclusions are summarized in the last
section.  

\section{The continuum limit}

We consider an anharmonic chain of particles with mass $M$ and
nearest-neighbor interactions. The particles interact via an anharmonic
potential with a cubic 
anharmonicity. The Hamiltonian of this system reads
\begin{eqnarray}\label{hamiltonian} 
&&H=\sum_n\left\{ \frac{P_n^2}{2 M} +\right.\nonumber\\
&&\left.G\left( \frac{1}{2}(Y_{n+1}-Y_n)^2+\frac{A}{3}(Y_{n+1}-Y_n)^3\right)\right\},\quad
\end{eqnarray} 
where $Y_n$ denotes the longitudinal displacement of the $n$-th
particle  from its equilibrium position, and
\begin{equation}\label{discretemom}
P_n = M\frac{dY_n}{dt} 
\end{equation}
is the momentum.
Here $G$ and $A$ are the potential parameters whose values depend
on the lattice. The associated first order equations of motion read
\begin{eqnarray} \label{motion0} 
\frac{dY_n}{dt}&=&\frac{1}{M}P_n\\
\label{motion00} 
\frac{dP_n}{dt}&=&-\frac{\partial H}{\partial Y_n}+F_n^{Noise}+F_n^{Damping},
\end{eqnarray}
where
\begin{eqnarray}\label{Hamilteq}
\frac{\partial H}{\partial
Y_n}&=&-G\left(Y_{n+1}-2Y_{n}+Y_{n-1}\right)\nonumber\\ 
&& -GA\left((Y_{n+1}-Y_{n})^2 - (Y_{n}-Y_{n-1})^2 \right).\quad\quad
\end{eqnarray}
In Eq. (\ref{motion00}) we have already  added  both  a stochastic force,
$F_n^{Noise}$, and a damping force,$F_n^{Damping}$. Both forces couple
the discrete system with a thermal bath. Here, we use the inner or
hydrodynamical damping, which reads \cite{Landau,arevalo}
\begin{equation}\label{damping1}
F_n^{Damping}=M\nu\left(\frac{dY_{n+1}}{dt}-2\frac{dY_n}{dt}+\frac{dY_{n-1}}{dt}\right).
\end{equation}
This means that the energy dissipation is provided by the 
irreversible processes arising 
from the finite velocity of the internal motions of the system, namely
time derivative of the relative displacements between particles in the
chain. Eq. (\ref{damping1}) is the discrete version
of the damping used in elasticity theory \cite{Landau}.
To fulfill the fluctuation-dissipation theorem the noise must have the
form (see App. {\bf A})
\begin{equation}\label{noiseespecial}
F_n^{Noise}=\sqrt{\rm D}\left(\xi_{n+1}(t)-\xi_{n}(t)\right)
\end{equation} 
where 
\begin{equation}\label{diffusion}
{\rm D}=2M\nu k_B T
\end{equation}
is the diffusion constant and $\nu$ is the damping constant.
$\xi_n(t)$ is
delta-correlated white noise, 
\begin{eqnarray}
\left\langle \xi_n(t)\xi_m(t^{'})\right\rangle &=&\delta_{nm}\delta(t-t^{'}),\\
\left\langle \xi_n(t)\right\rangle &=&0.
\end{eqnarray}
Since our interest is the study  of the lattice soliton diffusion close
to the sound velocity, $c$, we can use the continuum limit approach,
where $Y_n(t)\rightarrow y(x,t)$ and $\xi_n(t) \rightarrow \xi(x,t)$
with $x=n\,a$ and $a$ the equilibrium atomic spacing. In this limit
\cite{Pnevmatikos}, Eq. (\ref{motion0}) reduces to a form of noisy and
damped KdV equation (see App. {\bf B} for details)
\begin{equation}\label{kdv0}
\partial_{\tau} u+6u\partial_s u+\partial_s^3 u
=\nu_1 \partial_{ss}u-\sqrt{D_1}\partial_s \overline{\xi}(s,\tau)
\end{equation}
where 
\begin{equation}\label{transform1}
s=\alpha (x-c t), \quad\quad\quad \tau= \beta t, \quad\quad\quad 
u=\gamma \partial_s y. 
\end{equation}
The constants $\alpha$, $\beta$ and $\gamma$ are defined in
(\ref{galileo2}), and
$\nu_1$ and  $D_1$ are given by Eqs. (\ref{kdv01}).
Note that 
\begin{eqnarray}\label{conditionnoise}
\left\langle \overline{\xi}(s,\tau)\overline{\xi}(s^{\prime},\tau^{\prime})\right\rangle 
=\delta(s-s^{\prime})\delta(\tau-\tau^{\prime}).
\end{eqnarray}
Here and in the following the line over $\overline{\xi}$ is omitted.

The  case $D_1=0$ reduces
Eq. (\ref{kdv0}) to the KdV-Burgers equation. The associated
KdV  equation is
\begin{equation}\label{kdv1}
\partial_{\tau} u+6u\partial_s u+\partial_s^3 u=0
\end{equation}
whose one-soliton solution reads
\begin{equation}\label{kdv2}
u_0(s,\tau)=2\eta^2_0\, \sech^2[\eta_0 (s-4\eta^2_0\tau-s_0)].
\end{equation}
Here
\begin{equation}\label{kdv3}
\eta_0 =\frac{1}{p}\sqrt{3c(v-c)}
\end{equation}
is the inverse soliton width and $s_0$ is the
initial soliton position. The sound velocity $c$ and the constant
$p$ are defined in (\ref{constants1}).

\section{Collective coordinate approach \label{seccollec}}

To analyze our problem we assume that the soliton profile,
$u_0(s,\tau)$, is not  disturbed by the noise and damping terms 
and that only the width and amplitude are modified.
This assumption is well satisfied for low-energy solitons, whose
velocity is close to the sound velocity, because tails induced by the
perturbations are small in this velocity regime \cite{grim,arevalo}. So
we introduce a generalized traveling wave ansatz of the form
\begin{eqnarray}\label{ansatz1}
u(s,\tau)&=&u_0(s-S(\tau),\eta(\tau))\nonumber\\
&=&2\eta^2(\tau)\, \sech^2[\eta(\tau) (s-S(\tau))],
\end{eqnarray}
where the collective variables $S(\tau)$ and $\eta(\tau)$ are  the
soliton position  and the inverse soliton width, respectively.
Here and in the following the index of the one-soliton solution $u_0$
is omitted.

To obtain the equations for our collective coordinates we follow
\cite{niurkaPRE,Mertens1}. First, by substituting (\ref{ansatz1})
into Eq. (\ref{kdv0}) we get
\begin{equation}\label{collective1}
\phi_1\,{\dot S(\tau)}+\phi_2\,{\dot \eta(\tau)}=
\nu_1\partial_{ss}u-\sqrt{D_1}\partial_s \xi(s,\tau),
\end{equation}
where 
\begin{eqnarray}\label{modes1}
\phi_1(s,\tau)=\frac{\partial u}{\partial S}
\end{eqnarray}
and
\begin{eqnarray}\label{modes2}
\phi_2(s,\tau)=\frac{\partial u}{\partial \eta}.
\end{eqnarray}
Notice that the functions $\left\{\phi_i\right\}_{i=1,2}$
coincide with the adiabatic approximation (omitting secular terms in time) of
the discrete solutions of the linearized KdV equation
around the one-soliton solution (\ref{kdv2}) \cite{Mann}. 
We remark here  that our collective coordinate theory does not
take into account the contribution of the phonon modes (continuous
basis function solution of the linearized KdV). We discuss the
effect of noise-induced phonons in section \ref{secdiff}. The functions
$\left\{\phi_i\right\}_{i=1,2}$ are also orthogonal, so
the inner product $\int ds \,\phi_i(s,\tau)\,h(s,\tau)$ projects a
function $h$ onto the functions $\left\{\phi_i\right\}_{i=1,2}$. 
Then, by projecting Eq. (\ref{collective1}) we get
\begin{equation}\label{collective2}
A_i{\dot S}(\tau)+B_i{\dot
\eta}(\tau)=f_i+f_i^{damping}+f_i^{Noise}\quad\quad i=1,2 \quad ,
\end{equation}
where
\begin{eqnarray}
A_i&=&\int ds \frac{\partial u}{\partial S}\,\phi_i,\\
B_i&=&\int ds \frac{\partial u}{\partial \eta}\,\phi_i,\\
f_i&=&\int ds \left(6u\partial_s u+\partial_s^3 u\right)\phi_i,\\
f_i^{damping}&=&\nu_1\int ds \partial_{ss}u\,\phi_i,\\
f_i^{Noise}&=&-\sqrt{D_1}\int ds \partial_s \xi(s,\tau)\,\phi_i.
\end{eqnarray}
After some calculations the Eqs. (\ref{collective2}) take the form
\begin{eqnarray}\label{collective3}
&&\frac{dS(\tau)}{d\tau}=4\eta^2(\tau)
+\frac{15\sqrt{D_1}}{64\eta^5(\tau)}\int ds(\partial_s \phi_1)\xi(s,\tau),\quad\quad\\
\label{collective3a}
&&\frac{d\eta(\tau)}{d\tau}=-\frac{30\nu_1}{30+\pi^2}\eta^3(\tau)+\nonumber\\
&&\quad\quad\quad
\frac{45\sqrt{D_1}}{16(30+\pi^2)\eta(\tau)}
\int ds (\partial_s \phi_2)\xi(s,\tau).
\end{eqnarray}
To achieve the calculations we have assumed that the soliton profile
remains mostly unaffected and only its width and amplitude
change due to the stochastic perturbations. Then, at least for  small noise,
we can perform the calculations by taking the soliton field out of the
averages. Moreover, we have interpreted Eqs. (\ref{collective3}) and
(\ref{collective3a}) in the Stratonovich sense, because it assumes
$\xi(s,\tau)$ is a real noise with finite correlation time, which is
then allowed to become infinitesimally small after calculating
measurable quantities \cite{Gardiner}. Notice that white noise means
taking the limit of zero correlation time.

Eqs. (\ref{collective3}) and (\ref{collective3a}) can take the form
\begin{eqnarray}\label{FPec1}
\frac{d{\bf Y}(\tau)}{d\tau}=
{\bf A}^{Str}[{\bf Y}(\tau)]+\int ds \hat{\bf B}^{Str}[s,{\bf
Y}(\tau)]\mbox{\boldmath $ \xi $}(s,\tau),
\end{eqnarray}
where the elements of the noise vector \mbox{\boldmath $ \xi$}
satisfy (\ref{conditionnoise}).
$\{Y_1,Y_2\}=\{S,\eta\}$ are the elements of the vector ${\bf Y}$, the
elements $\{A^{Str}_1,A^{Str}_2\}$ of the drift vector ${\bf A}^{Str}$
are the drift terms in Eqs. (\ref{collective3}) and
(\ref{collective3a}), respectively. The diffusion matrix $\hat{\bf
B}^{Str}$ is diagonal, where $B^{Str}_{11}$ and $B^{Str}_{22}$ are the 
coefficients in front of the noise in Eqs. (\ref{collective3}) and (\ref{collective3a}),
respectively. In order to facilitate the calculations we write
Eq. (\ref{FPec1}) in the Ito form,
\begin{eqnarray}\label{FPec2}
d{\bf Y}(\tau)={\bf A}^{Ito}[{\bf Y}(\tau)]d\tau+
\int ds \hat{\bf B}^{Ito}[s,{\bf Y}(\tau)]d{\bf W}(s,\tau),\nonumber\\
\end{eqnarray}
where the $d{\bf W}(s,\tau)=\mbox{\boldmath $ \xi $}(s,\tau)d\tau$
is a Wiener process. Via a Fokker-Planck equation, one can show that
the elements of the drift vector ${\bf A}^{Ito}$  read \cite{Gardiner}
\begin{eqnarray}
&&A^{Ito}_i[{\bf Y}(\tau)]=A^{Str}_i[{\bf Y}(\tau)]+\nonumber\\
&&\quad\quad\quad\frac{1}{2}\sum_{j\,k\,m}\int ds
B^{Str}_{k\,m}[s,{\bf Y}(\tau)]\partial_{Y_k}B^{Str}_{i\,j}[s,{\bf
Y}(\tau)]\nonumber\\
&&\quad\quad\quad i,j,k=1,2
\end{eqnarray}
while
\begin{eqnarray}
\hat{\bf B}^{Ito}[s,{\bf Y}(\tau)]=\hat{\bf B}^{Str}[s,{\bf Y}(\tau)].
\end{eqnarray}
Notice that ${\bf A}^{Ito}$ and $\hat{\bf B}^{Ito}$ are
nonanticipating functions. So,
from Eq.(\ref{FPec2}) it is easy to show the following averages
\begin{eqnarray}\label{collective4}
\left\langle S(\tau)\right\rangle  &=& 
\langle \int\limits_{0}^{\tau} d\tau^{\prime} 4\eta^2(\tau^{\prime})\rangle \nonumber\\
\left\langle \eta(\tau)\right\rangle &=&
-\bigg\langle \int\limits_{0}^{\tau} 
d\tau^{\prime}\frac{30\nu_1}{30+\pi^2}\eta^3(\tau^{\prime})\bigg\rangle +\nonumber\\
&& \int\limits_{0}^{\tau} d\tau^{\prime}\frac{225(231+8\pi^2 )D_1}{112(30+\pi^2)^2} \nonumber\\
Var\left(S(\tau)\right)&=&
\bigg\langle \int\limits_{0}^{\tau} d\tau^{\prime}
\frac{75\,D_1}{112\,\eta^3(\tau^{\prime})}\bigg\rangle \nonumber\\
Var\left(\eta(\tau)\right)&=&\bigg\langle \int\limits_{0}^{\tau}
d\tau^{\prime}\frac{225(21+\pi^2)\,D_1\eta(\tau^{\prime})}{28(30+\pi^2)^2}\bigg\rangle
\nonumber\\ 
Corr\left(S(\tau)\eta(\tau)\right)&=&0.
\end{eqnarray}
Here $\langle \,\cdot\cdot\cdot\,\rangle $ means average over an ensemble of
realizations, 
\begin{equation}
Corr(PQ)=
\langle \,PQ\,\rangle -\langle \,P\,\rangle \langle \,Q\,\rangle \quad
\textrm{and}\quad Var(P)
=Corr(PP).
\end{equation}
Now we define a new set of Langevin equations, 
\begin{eqnarray}\label{collective5}
&&dY_i(\tau)=a_id\tau+\sum_jb_{ij}\,dW_j(\tau)\nonumber\\ 
&&\textrm{with}\quad i,j=1,2 \quad \textrm{and} \quad \{Y_1,Y_2\}= \{S,\eta\},
\end{eqnarray} 
which we have interpreted in the Ito sense. 
$dW_j(\tau)=\xi_j(\tau)d\tau$ are Wiener processes where
we have let the noises to be
uncorrelated, namely
\begin{eqnarray}
\langle \xi_j(\tau)\xi_{j^{\prime}}(\tau^{\prime})\rangle
=\delta(\tau-\tau^{\prime})\delta_{jj^{\prime}}.
\end{eqnarray}
In order to determine the values of $a_i$ and $b_{ij}$
we have demanded that Eqs.(\ref{collective5}) satisfy the
relations (\ref{collective4}). It is 
straightforward to see that Eqs. (\ref{collective5}) take the form 
\begin{eqnarray}\label{collective6}
dS(\tau)&=&4\eta^2(\tau)d\tau+
\frac{5\sqrt{3}}{4\sqrt{7}}\sqrt{\frac{D_1}{\eta^3(\tau)}}\,dW_1(\tau)\\
\label{collective7}
d\eta(\tau)&=&\bigg(-\frac{30\nu_1}{30+\pi^2}\eta^3(\tau)
+\frac{225(231+8\pi^2 )D_1}{112(30+\pi^2)^2}\bigg)d\tau+\nonumber\\
&&\frac{15\sqrt{21+\pi^2}}{2\sqrt{7}(30+\pi^2)}\sqrt{D_1\eta(\tau)}\,dW_2(\tau).
\end{eqnarray}
Eqs. (\ref{collective6}) and (\ref{collective7}) are statistically
equivalent to  Eqs (\ref{collective3}) and (\ref{collective3a})
because they share the same  Fokker-Planck equation. 
Since the derivation of Eqs (\ref{collective6}) and
(\ref{collective7})  involved approximations, we have not solved them
exactly. Instead of that, we have used perturbation analysis \cite{Abdullaev},
which is developed in detail in  App. {\bf C}. In order to do so,
we have considered the thermal terms as perturbations, so Eqs. (\ref{collective6}) and
(\ref{collective7}) take the form
\begin{eqnarray}\label{collective8}
dS(\tau)&=&4\eta^2(\tau)d\tau+\epsilon
\frac{5\sqrt{3}}{4\sqrt{7}}\sqrt{\frac{D_1}{\eta^3(\tau)}}\,dW_1(\tau)\\
\label{collective9}
d\eta(\tau)&=&-\frac{30\nu_1}{30+\pi^2}\eta^3(\tau)d\tau+
\epsilon\bigg(
\frac{225(231+8\pi^2 )D_1}{112(30+\pi^2)^2}d\tau+\nonumber\\
&&\frac{15\sqrt{21+\pi^2}}{2\sqrt{7}(30+\pi^2)}\sqrt{D_1\eta(\tau)}\,dW_2(\tau)\bigg).
\end{eqnarray}
Now, we seek an asymptotic solution in the form of a small-noise expansion 
\begin{eqnarray}\label{pertur1}
S(\tau)=s_0(\tau)+\epsilon\,s_1(\tau)+\cdot\cdot\cdot\nonumber\\
\eta(\tau)=\eta_0(\tau)+\epsilon\,\eta_1(\tau)+\cdot\cdot\cdot.
\end{eqnarray}
Here, $\epsilon$ is a factor introduced for
convenience in the analytical calculations. Notice that the case
$\epsilon=0$ reduces Eqs. (\ref{collective8}) and (\ref{collective9})
to the deterministic case. In order to interpret our  perturbation
theory we must set $\epsilon=1$ and assume that the terms on the
r.h.s. of Eqs. (\ref{collective8}) and (\ref{collective9}) are
sufficiently small. So we must restrict ourselves to a regime of low temperatures
of the thermal bath ($D_1$ small). 
From the perturbation analysis we obtain that 
\begin{widetext}
\begin{eqnarray}\label{pertur2}
\langle S(\tau) \rangle &=&
\langle s_0(\tau)\rangle+\langle s_1(\tau) \rangle\nonumber\\ 
 &=&4\frac{\eta_0^2(0)}{\lambda}\log{(1+\lambda\tau)}
+ \frac{15D_1\,(231+8\pi^2)\,\eta_0(0)\,(2(1+\lambda\tau)^{5/2}-5\lambda\tau-2)}
{7(30+\pi^2)^2\lambda^2(1+\lambda\tau)}, \nonumber\\ 
  \nonumber\\ 
\langle 4\,\eta^2(\tau) \rangle &=&
\langle4\,\eta_0^2(\tau)+8\eta_0(\tau)\eta_1(\tau)\rangle\nonumber\\ 
 &=&\frac{4\,\eta_0^2(0)}{1+\lambda\tau}
+\frac{45D_1(231+8\pi^2)\,\eta_0(0)\,((1+\lambda\tau)^{5/2}-1)}
{7(30+\pi^2)^2\lambda\,(1+\lambda\tau)^2},\nonumber\\
  \nonumber\\ 
Var(S(\tau))&=&D_1\bigg(
\frac{-15}{56\,\eta^3_0(0)\,\lambda } - 
\frac{480\,(21 + {\pi }^2)\eta^3_0(0)(8+7\lambda\tau(5\lambda\tau+4))}{49\,{\left( 30 + {\pi }^2 \right) }^2\,{\lambda }^3\,
{\left( 1 + \lambda\,\tau  \right) }^2} \nonumber\\
&+& \frac{15\,(1 + \tau \,\lambda)^{3/2} }
{392\,{\left( 30 + {\pi }^2 \right) }^2\,\eta^3_0(0)\,{\lambda }^3}
\left(2048(21+\pi^2)\eta^6_0(0)+7(30+\pi^2)^2\lambda^2(1+\lambda\tau)\right)\bigg),\nonumber\\
  \nonumber\\  
Var(4\,\eta^2(\tau)) &=&\frac{7200 D_1
(21+\pi^2)\eta^3_0(0)}{49\lambda (30+\pi^2)^2}
\bigg(\frac{1}{\sqrt{1+\lambda\tau}}-\frac{1}{(1+\lambda\tau)^4}
\bigg).\nonumber\\
\end{eqnarray}
\end{widetext}
The  expressions for $D_1$ and $\lambda$ are given by Eqs. (\ref{kdv01}) and
(\ref{appB81}), respectively.

\section{Simulations \label{simulations}}

Substituting Eqs. (\ref{damping1}) and (\ref{noiseespecial}) into
Eq. (\ref{motion0}) we get the full set of discrete equations of
motion written in absolute displacements. However, for our simulations
relative displacements are more convenient, because the
 lattice solitons in this representation are
pulse solitons whose amplitude  vanishes at
infinity. This characteristic allows us  to  use periodic
boundary conditions which are necessary for long simulation times,
because we want to avoid reflections at the boundaries.
So the discrete equations of motion in relative displacements read 
\begin{eqnarray}\label{simul1}
& &M \frac{d^2V_n}{dt^2}=G\left(V_{n+1}-2V_{n}+V_{n-1} \right)+\nonumber\\
& &\quad\quad GA\left(V_{n+1}^2 - 2V_{n}^2+ V_{n-1}^2 \right)+\nonumber\\
& &\quad\quad  M\nu\left(\frac{dV_{n+1}}{dt}-2\frac{dV_n}{dt}
+\frac{dV_{n-1}}{dt}\right)+ \nonumber\\
& &\quad\quad \sqrt{\rm D}\left(\xi_{n+1}(t)-2\xi_{n}(t)+\xi_{n-1}(t)\right),
\end{eqnarray}
where $V_n(t)=Y_{n+1}(t)-Y_{n}(t)$ and ${\rm D}=2M\nu k_B T$.
The periodic boundary conditions read
\begin{eqnarray}
\frac{d^l\,V_0}{dt^l}=\frac{d^l\,V_{N-1}}{dt^l}, 
&\quad& 
\frac{d^l\,V_{N}}{dt^l} = \frac{d^l\,V_1}{dt^l}, \quad\quad
l=0,1\nonumber\\
\nonumber\\
\xi_0(t)=\xi_{N-1}(t), &\quad & \xi_{N}(t)=\xi_1(t),
\end{eqnarray}
where $N$ is the number of particles of our chain and $N-1$
is the number of bonds.

A suitable method to detect the position of a
pulse lattice soliton, $V_n(t)$, is to search for its 
maximum \cite{arevalo}. However, in the presence of
stochastic perturbations this method is not useful since the
pulse shape is strongly masked by the noise, an example of this
situation is shown in App. \ref{profiles}. So from the data of our
simulations we have taken snapshots of the system at different times,
and from them we 
have generated the kink shape $Y_n(t)$ of the lattice soliton by using
the algorithm
\begin{equation}\label{algorithm}
Y_n(t)=Y_1(t)+\sum_{i=1}^{n-1}V_i(t)\quad n=2,3,\cdot\cdot\cdot,N.
\end{equation} 
The kink shape is less distorted by the noise than the pulse shape
$V_i(t)$. In (\ref{algorithm}) $Y_1(t)$ is a boundary
condition that we have demanded to be  
\begin{equation}\label{Y1}
Y_1(t)=-\frac{1}{2}\sum_{i=1}^{N-1}V_i(t),
\end{equation}
so at $t=0$ the amplitude of the center of the kink shape is zero, as
it should be from the theory \cite{Pnevmatikos}.
Notice that
\begin{equation}\label{conservq}
Y_{N}(t)-Y_1(t)=\sum_{i=1}^{N-1}V_i(t)
\end{equation}
is a conserved quantity in our system, i.e.
\begin{equation}\label{conservq1}
{\dot Y}_{N}(t)-{\dot Y}_1(t)=\sum_{i=1}^{N-1}{\dot V}_i(t)=0.
\end{equation}
We have checked Eq. (\ref{conservq1})  with a precision higher than $10^{-14}$
over the whole time range of our Langevin dynamics simulations.

In order to determine every time the parameters of the soliton, namely soliton
velocity $v$ and position $x$, we have proceeded as follows. 
We have searched for the values of the parameters $x$ and $v$ where the
relation 
\begin{eqnarray}\label{algoritm}
\sum\limits_{n=n_0-n_1}^{n_0+n_1} \left(Y_n-y_0(n\,a-x,v)\right)=0
\end{eqnarray}
is fulfilled. Here
\begin{equation}\label{restframe1}
y_0(n\,a-x,v)=\frac{6\sqrt{2\,h\,c (v-c)}}{p }
\tanh\left(\frac{n\,a-x}{L(v)}\right) 
\end{equation} 
with
\begin{equation}\label{restframe2}
L(v)\,=\,(\alpha\,\eta(t))^{-1}\,=\,2\sqrt{\frac{h}{2\,c\,(v-c)}}.
\end{equation}
Here the function $\eta$ is defined in (\ref{kdv3}) and the constants
$h$, $p$, $c$ and $\alpha$ are defined in (\ref{constants1}) and
(\ref{galileo2}). the function (\ref{restframe1}) is the absolute
displacement representation of the one-soliton solution (\ref{kdv2})
in a frame moving with the soliton velocity. 
In Eq. (\ref{algoritm}) 
\begin{equation}\label{defs}
n_0=\textrm{int}(x)\quad \textrm{and}\quad n_1=\textrm{int}(\frac{3}{2} L(v(0))),
\end{equation}
where $\textrm{int}(\cdot)$ denotes the integer part of a number and
$v(0)$ is the initial soliton velocity. The value
of $n_1$ has been chosen to take into account only the core of the
lattice kink-shape and it is constant during our simulations.
In order to determine both $x$ and $v$ we consider different values
of  $v$ in Eq. (\ref{algoritm})
within a range of velocities around the initial soliton velocity,
namely $v-c\, \in\, [0.1\,(v(0)-c),2\,(v(0)-c)]$. For every value of
$v$ we search the value $x$ that fulfills Eq. (\ref{algoritm}), so
we get a set of pairs of values $x$ and $v$. 
Finally, from this set of pairs of values we
search, by using linear interpolation,  the values of $x$ and $v$
which fulfill the relation
\begin{eqnarray}\label{algoritm1}
\frac{\sum\limits_{n=n_0-n_1}^{n_0+n_1}Y_n\,y_0(n\,a-x,v)}
{\sum\limits_{n=n_0-n_1}^{n_0+n_1}(y_0(n\,a-x,v))^2}=1\quad.
\end{eqnarray}
Notice that in Eqs. (\ref{algoritm}) and (\ref{algoritm1}) we have
assumed that the lattice kink shape, $Y_n$, is closely related with the
function $y_0$, however, as was mentioned in Ref. \cite{arevalo},  a pulse
lattice soliton in the presence of damping develops a tail. The
amplitude of this trailing tail depends on both soliton the velocity and
the damping, so it is bigger when the damping and/or the soliton velocity is
higher. Thus, we  restrict ourselves to velocities very
close to the sound velocity where the effect of this trailing tail
is negligible. 

Up to now we have determined the parameters $x$ and $v$, which fit the
function $y_0$ to the lattice kink shape $Y_n$, so we have not
measured directly either $x$ or $v$. Since the function $y_0$ is closely
related to the lattice kink-shape $Y_n$, one could assume both  $x$ and
$v$ as an estimate of the soliton position and  velocity,
respectively. However, we have taken only the parameter $v$ as an
estimate of the soliton velocity and with this value we have used a different
method to determine the soliton position. In fact, in order 
to be in agreement with our collective coordinate approach,
where we have projected the equations of motion onto the Goldstone mode $\phi_1$
(Eq. (\ref{modes1})), we have projected the noisy
kink shape $Y_{n}(t)$ onto the pulse solution $u_0$
defined in (\ref{kdv2}). Notice that in the absolute displacement
representation the function $u_0$ is the Goldstone mode. So this
projection reads
\begin{equation}\label{project1}                
P(x)=\sum_{i=n-n_2}^{n+n_2} Y_{i}(t)u_0(i\,a-x,v)
\end{equation}
where
\begin{equation}
u_0(i\,a-x,v)=\frac{6c(v-c)}{p}\sech^2\left(\frac{i\,a-x}{L(v)}\right)
\end{equation}
and $x=n\,a$.
The value of $n_2$ in Eq. (\ref{project1}) is much larger than the
soliton width, so the boundary 
effects are negligible. The function $u_0(i\,a-x,v)$ is the
one-soliton solution (\ref{kdv2}) in a frame moving with the soliton
velocity. Afterwards
we have searched, by linear interpolation, the value $x$ where $P(x)$
vanishes and we have defined it as the position of the soliton center
of mass. At this point we remark that the values of $x$ following from
this latter method are not significantly different from the 
values of $x$ following from the former method
(Eqs. \ref{algoritm} and \ref{algoritm1}). However, we consider  the
latter method to be more appropriate than the former one in the sense
that we proceed in our code in a similar way as in our
analytical calculations. 

Our Langevin dynamics simulations were performed for a chain with 1500
lattice points. The time integration was carried
out by using the Heun method \cite{Heun}, which has been successfully
used in the numerical solution of partial differential equations and
difference-differential equations, coupled to
either an additive or a multiplicative noise term
\cite{kampetter,niurka1,mathias1,mathias}. 
Here, we have used the conserved quantity (\ref{conservq})
to check the accuracy of our code \cite{arevalo}. For the
longest  simulation time the variation of this conserved quantity has
been lower than  $10^{-9}\,\%$. In order to start the
simulations at $t=0$ we have used the  one-soliton solution
(\ref{kdv2}) of the KdV equation in the laboratory frame. 
The average values have been calculated over
200 realizations up to a final time 5000.
All values of the constants of the equation (\ref{simul1}) are
set at unity except the damping constant
which is set at $\nu=0.003$. Notice that for lower values of damping
the relaxation of the system energy would take more time in our
simulations to reach a regime close to its stationary value.
On the other hand, higher values of damping can
strongly distort the soliton shape, namely the tail induced by the
damping cannot be neglected when the value of the damping is high. In
App. \ref{thermalization} we show the thermalization process in our
system. In our simulations the values of temperature,
$T$, and initial soliton velocity, $v(0)$, are parameters (see figure
captions).  
\begin{figure}
\includegraphics{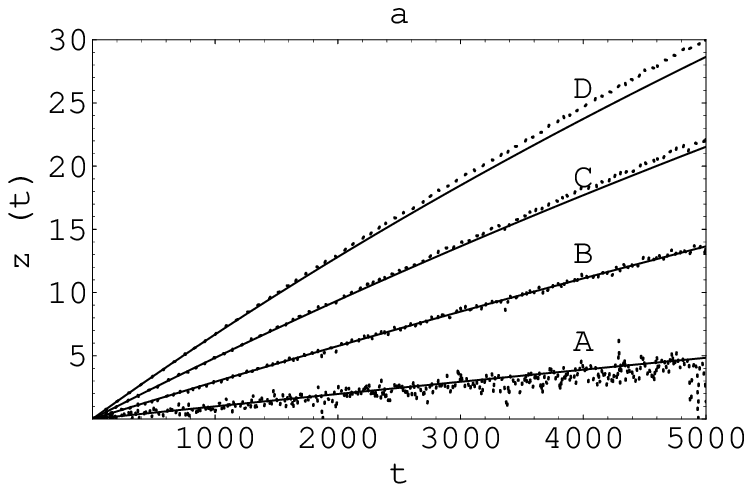}
\includegraphics{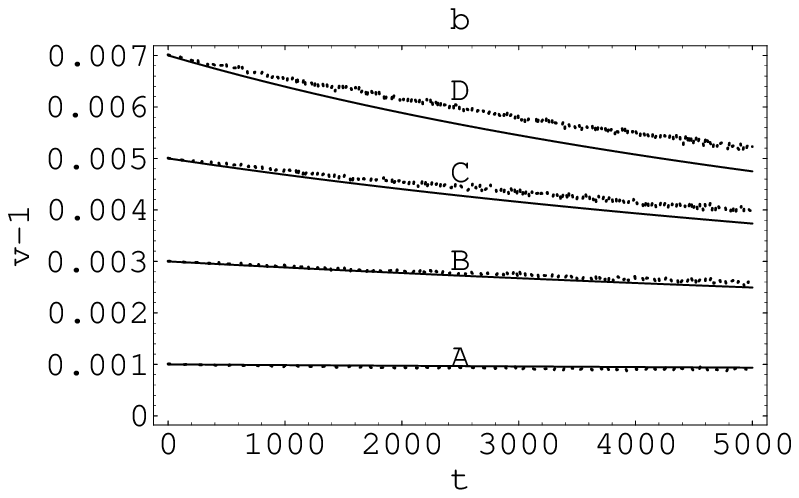}
\caption{Averaged soliton position (a) and velocity (b) vs. time
in the sound velocity frame $z=x-t$ ($c=1$), with
$\nu\,=\,0.003$ and $T=5\times 10^{-5}$. Dotted lines: simulation, 
solid lines:  theory (Eqs.(\ref{pertur2})). A, 
B, C, D, and E correspond to different initial velocities,  
namely $\,v(0)=1.001$, $1.003$, $1.005$ and $1.007$,
respectively.}
\label{fig1}
\end{figure}
\begin{figure*}
\includegraphics{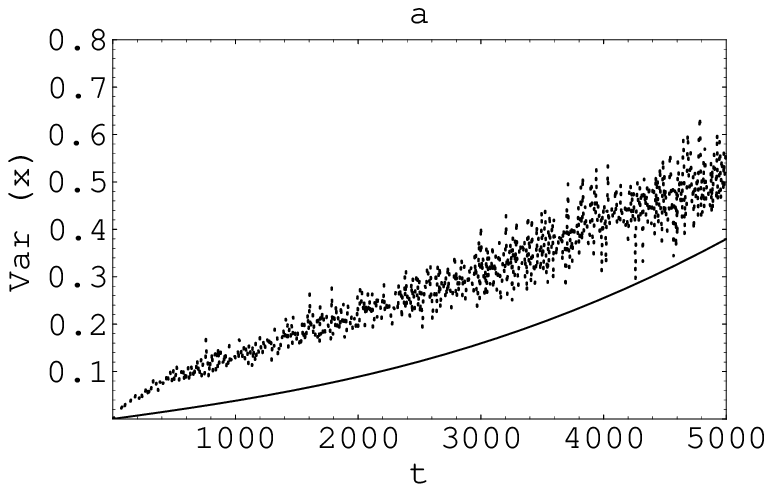}
\includegraphics{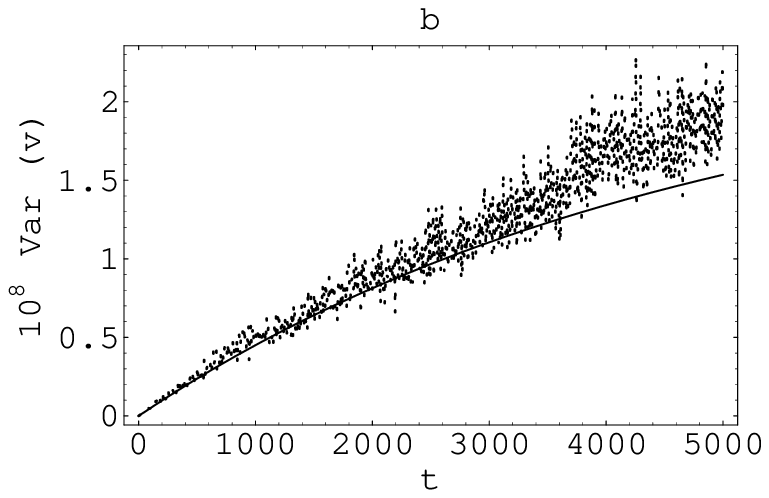}
\includegraphics{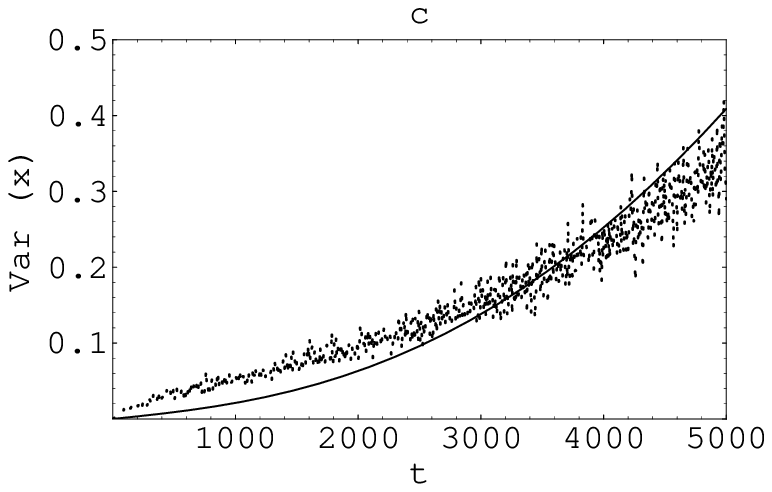}
\includegraphics{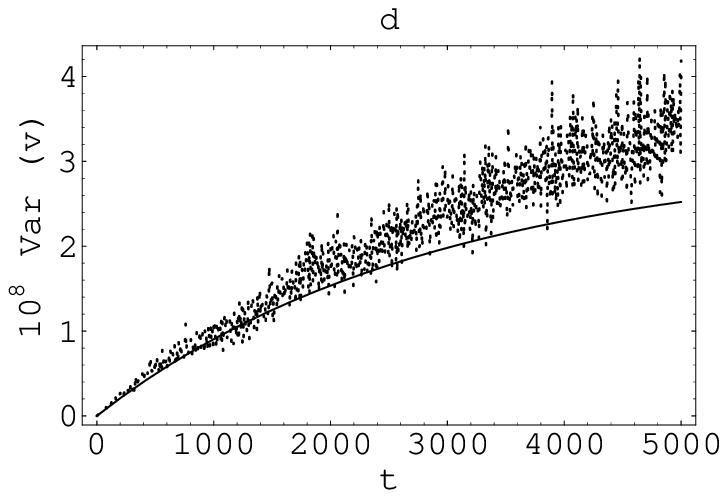}
\includegraphics{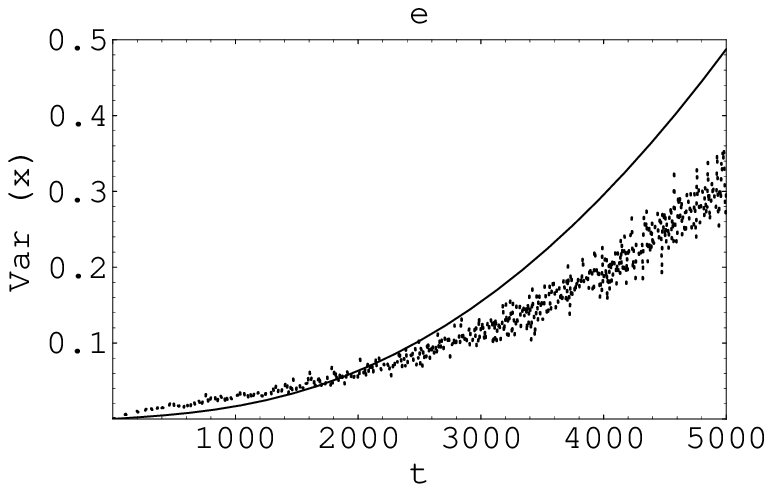}
\includegraphics{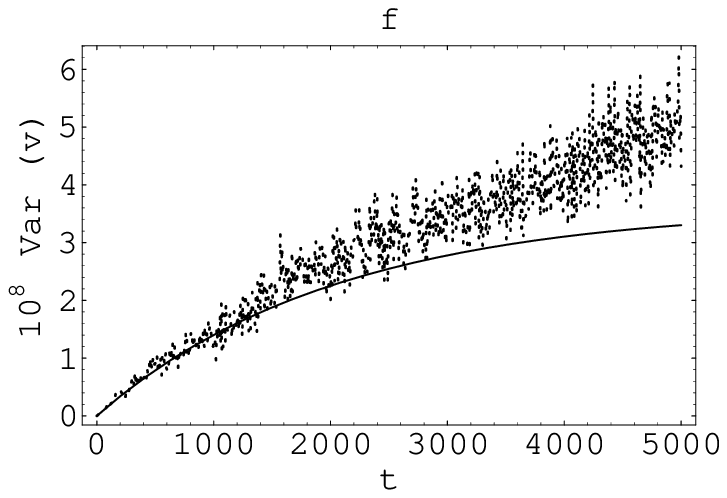}
\caption{Variances of the soliton position (panels a, c and e) and velocity
(panels b, d and f) of the soliton vs. time, 
with $\nu\,=\,0.003$ and $T=5\times10^{-6}$. Dotted lines: simulation,
solid lines: theory (Eqs.(\ref{appB11}) and (\ref{appB12})). 
The panels correspond to different initial velocities,  
namely $\,v(0)=1.003$ (a and b), $1.005$ (c and d) and $1.007$
(e and f). The reduced temperatures are $\overline{T}=0.0061419$,
$0.00283324$, $0.00169765$, respectively.}  
\label{fig3}
\end{figure*}
\begin{figure*}
\includegraphics{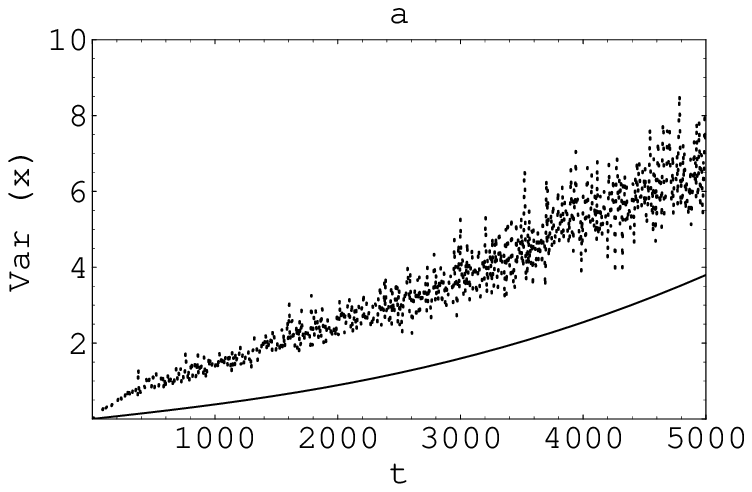}
\includegraphics{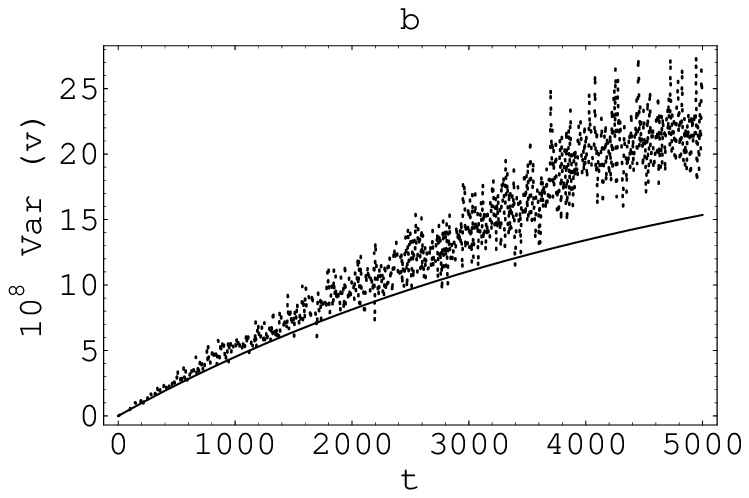}
\includegraphics{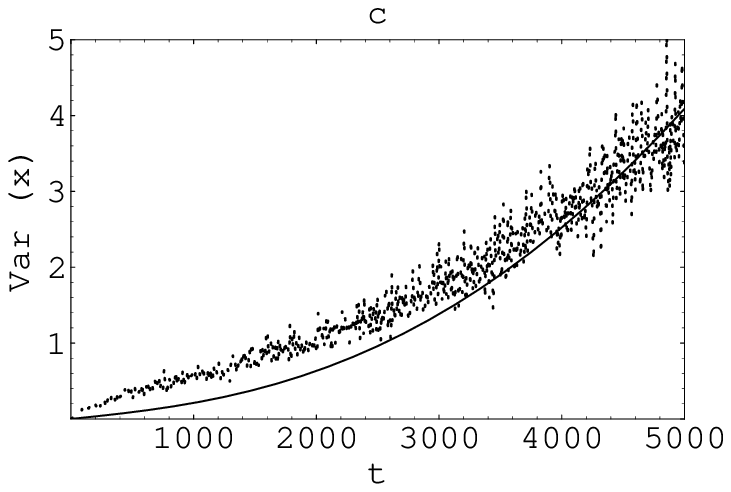}
\includegraphics{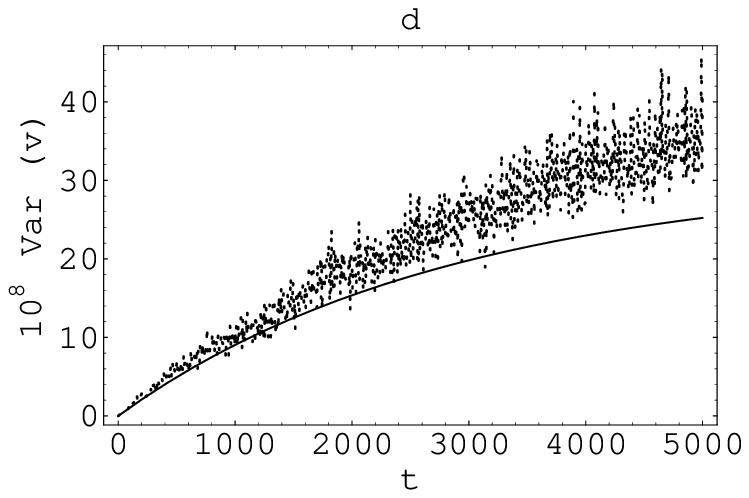}
\includegraphics{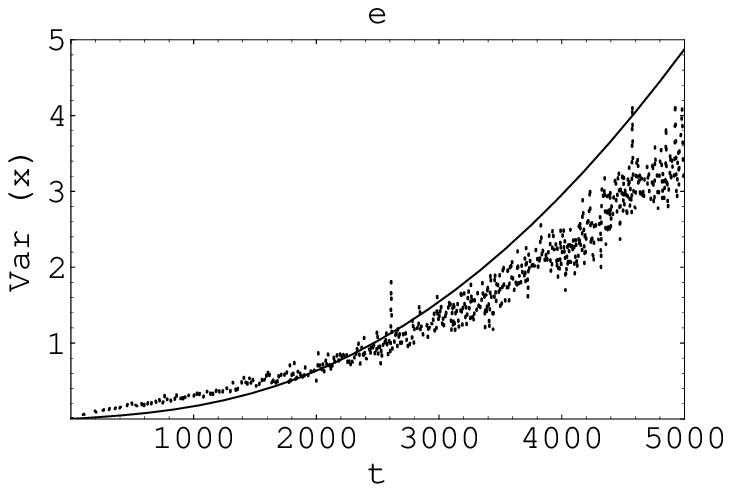}
\includegraphics{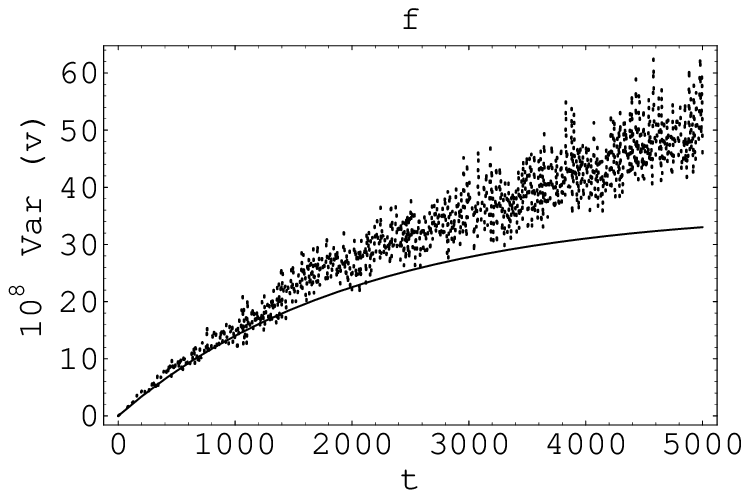}
\caption{Variances of the soliton position (panels a, c and e) and velocity
(panels b, d and f) of the soliton vs. time,
with $\nu\,=\,0.003$ and $T=5\times10^{-5}$. Dotted lines: simulation,
solid lines: theory (Eqs.(\ref{appB11}) and (\ref{appB12})). 
The panels correspond to different initial velocities,  
namely $\,v(0)=1.003$ (a and b), $1.005$ (c and d) and $1.007$
(e and f). The reduced temperatures are $\overline{T}=0.061419$,
$0.0283324$, $0.0169765$, respectively.} 
\label{fig4}
\end{figure*}

\subsection{Soliton propagation}

In Figs. \ref{fig1}a and \ref{fig1}b we show several examples of both the 
averaged soliton position, $\langle x(t)\rangle$, and the averaged
soliton velocity, $\langle v(t)\rangle$, as functions of time from
both the simulation  and the theory (see Eqs.(\ref{pertur2})). Notice 
that 
\begin{eqnarray}
\langle x(t)\rangle=\frac{1}{\alpha}\langle S(\beta\,t) \rangle
+c\,t 
\end{eqnarray}
and
\begin{eqnarray}
\langle v(t)-c\rangle=\left(\frac{p^2}{12\,c}\right)
\langle 4\,\eta^2(\beta\,t)\rangle.
\end{eqnarray}
For all cases the soliton position from the
simulation agrees well with the position given by the
analytical theory. In the case of the soliton velocity, the agreement is better for
initial velocities close to the sound velocity
than for higher velocities. In
fact, the time evolution of the velocity given by the simulation is
always higher than the theoretical prediction. This systematic difference may
be due to the small tail that is generated by the lattice soliton
since it is a non-topological soliton \cite{arevalo}. So it may affect our numerical
method for determining the soliton position. We point out that the 
amplitude of this tail depends on both the soliton velocity and
the damping constant, and can be neglected only for velocities close to the
sound velocity and small values of the damping constant
\cite{arevalo}. This is the most important reason for restricting
our study to low-energy solitons whose
velocities are close to the sound velocity. Since this
difference is systematic it does not play any role in the 
numerical calculation of the variances which is our more important
goal in this article. 

\subsection{Soliton diffusion \label{secdiff}}

In Figs. \ref{fig3} and \ref{fig4} we show the variances of the
soliton position and velocity vs. time for different initial
velocities. The temperature of the thermal bath  in
Fig. \ref{fig4} is 10 times 
higher than that in Fig. \ref{fig3}. 
The results scale very well by a factor of 10; i.e. the
variances are proportional to the temperature, as expected from
Eqs. (\ref{pertur2}). 

Notice that our theory
(solid lines) has no adjustable parameters. Taking into account that
the theory consists of several steps (discrete system $\rightarrow$ Bq equation
$\rightarrow$ KdV equation $\rightarrow$ collective coordinate theory
$\rightarrow$ perturbation analysis) it is already a significant success to
obtain the observed results, as seen in Figs. \ref{fig3} and
\ref{fig4} 
 
We observe that the behavior of the  variances depends strongly on the initial
soliton velocity. For low-energy solitons, whose velocities are 
close to the sound velocity (Figs. \ref{fig3} and \ref{fig4}, cases (a)
and (b)), the soliton diffusion tends to be nearly normal, i.e. linear
in time. In fact, our theory predicts a normal diffusion for times
\begin{equation}\label{estimation}
t\,<<t^{*}=\,\left(\frac{30+\pi^2}{495}\right)\left(\frac{c}{\nu\,(v(0)-c)}\right).
\end{equation}
This estimate was obtained by comparing the first with second terms
of the Taylor expansion in powers of $\tau$ of the variance of the
velocity (see Eq. (\ref{appB13})). For low-energy solitons
($v(0){}^{>}_{\sim}\,c$) $t^{*}$ is much larger than our
simulation time (Figs. \ref{fig3}a and \ref{fig4}a). This means that
the superdiffusivity is not very dominant. However, for
higher-energy solitons the anomalous behavior 
turns out to be important after some time. In those cases
$t^{*}$ is comparable with our simulation time (Figs. \ref{fig3}e and
\ref{fig4}e).   

In the case of solitons with very low energy
the variance of the position (Figs. \ref{fig3}a and \ref{fig4}a) does
not agree so well  with the theoretical prediction. In fact, we
observe a transient behavior for times $t\,{}^{<}_{\sim}3/\nu=1000$
where the system energy shows a fast relaxation process (see
App. \ref{thermalization}). Those discrepancies between theory and
simulations may be due to the combination of two effects. First, 
the profile of low-energy solitons is strongly  masked by the noise, so
the numerical detection of the position can be distorted.
Second, not only the noise but also the
noise-induced phonons can make a significant contribution to the
variance of the position, since the reduced temperature,
$\overline{T}=k_B\,T/H(0)$ (temperature in units of the initial 
soliton energy $H(0)$), of the thermal bath is higher here than in the
other cases (see captions
Figs. \ref{fig3} and \ref{fig4}). In this
respect we estimate the phonon effect on the diffusion of
low-energy solitons in the next section (see also Fig. \ref{fig5}). 

Notice that the variance of the soliton position is
larger for low-energy solitons (Figs. \ref{fig3}a and \ref{fig4}a)
than for higher-energy solitons (for instance 
Figs. \ref{fig3}e and \ref{fig4}e). This is due to the fact
that the higher-energy solitons are
more robust against thermal
fluctuations than the lower-energy ones. Or, equivalently, the
reduced temperature $\overline{T}$ 
of the thermal bath is higher for slow solitons than for 
the fast ones (see captions of  Figs. \ref{fig3} and \ref{fig4}). 

On the other hand, the superdiffusive behavior is more pronounced for
higher-energy solitons. This is because  the soliton
velocity turns out to be more sensitive to the thermal fluctuations in this case 
than in the case of broader solitons. Notice that the  soliton velocity and
soliton width are related. Also, since the higher-energy solitons
encompass few lattice sites, the soliton-width perturbations are larger with
respect to the averaged soliton width in this case than in the case of broader solitons
(low-energy solitons). In fact, the variance of the soliton velocity
shows this effect, namely that for broader solitons (Figs. \ref{fig3}b and \ref{fig4}b)
this variance is smaller than for narrower solitons (for instance Figs. \ref{fig3}f and
\ref{fig4}f). The discrepancy between our theory and the numerical
simulations for higher-energy solitons (Figs. \ref{fig3} and
\ref{fig4}, cases (c)-(f)) is mainly due to the fact that our theory
is valid only for soliton velocities close to the sound velocity. 

With respect to the variance of the soliton velocity 
(Figs. \ref{fig3} and \ref{fig4}, panels b, d and e)
we observe that it is mostly anomalous and its behavior is nearly quantitatively 
predicted by our theory for $0\,\leq \,t\, {}^{<}_{\sim}\,2000$ in all
the cases. For larger times, $2000\,\leq \,t\, <\,5000$, there
is a discrepancy which becomes larger with  increasing initial velocities.

We remark that the numerical results shown in Figs. \ref{fig3} and
\ref{fig4} do not change for systems with the double number of
sites, namely 3000.

We comment that there was a previous attempt by
Scalerandi et al. \cite{ScalerandiRomano} to calculate theoretically
the mean square displacement of a KdV soliton subject to stochastic
fluctuations. They considered the case of small Stokes
damping and  a simple white noise delta-correlated in  time and space.
Though their theoretical result shows the appearance of
a noise-induced superdiffusive behavior, it does not have 
the same dependence with respect to
the soliton width as our results (\ref{pertur2}), which agree well
with our simulations.

\subsection{Estimate of the phonon contribution \label{estimated}}

\begin{figure}
\includegraphics{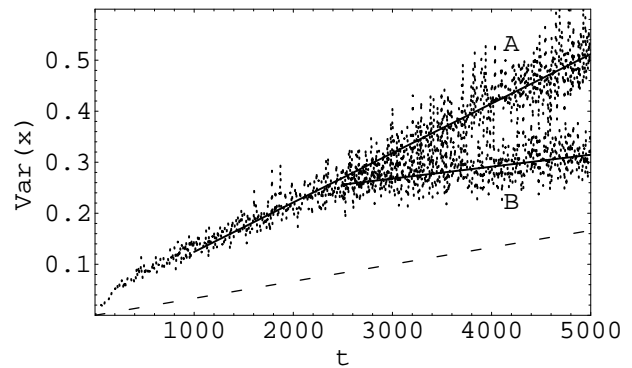}
\caption{An example of a test to determine the phonon contribution to
the diffusion constant. Comparison of the behavior of $Var(x)$ from
simulations (dotted lines) when the system is in contact (A) and
isolated (B: noise and damping are switched off for $t>2500$) from the
thermal bath. The slope of the straight lines (solid lines) fitted to
the simulation data (dotted) in both cases
give the observable values of
the diffusion constant, namely $D_{total}$ (A)  and $D_{ph}$ (B).
The difference can be compared with the slope $D_{th}$ of the dashed line
(linear part of the theory, Eq. (\ref{appB11})).
$v(0)=1.003$, $T=5\times10^{-6}$, $\overline{T}=0.0061419$.}
\label{fig5}
\end{figure}

\begin{figure}
\includegraphics{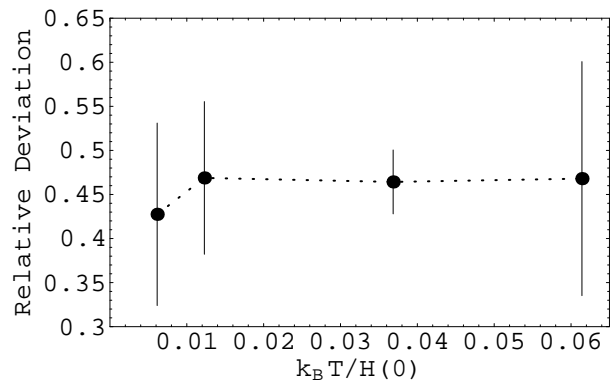}
\caption{Relative deviation $((D_{noise}-D_{th})/D_{noise})$ of
the diffusion coefficient vs. $k_B\,T/H(0)=\overline{T}$ with $v(0)=1.003$.} 
\label{fig6}
\end{figure}

In order to estimate this contribution we have performed the
following numerical test. We have simulated the propagation of
low-energy solitons under thermal fluctuations up to a time (e.g. 2500) 
when the system energy is close to its stationary value (see
App. \ref{thermalization}). Notice that the diffusion of low-energy
solitons is mostly normal (see Figs \ref{fig3}a and \ref{fig4}a).
Afterwards, we have isolated the system from the
thermal bath by switching off noise and damping, 
so that solitons propagate only in the noise-induced phonon bath. The
diffusion is mostly normal before and after the system is isolated,
i.e. the variance of the position is linear in time. We have compared
the slope of the variance of the isolated system ($t\,>\,2500$)  
with the slope of the variance in the case when the system is in
contact with the thermal bath the whole time. An example of this
test is shown in Fig. \ref{fig5}. We observe that both slopes, after
switching off noise and damping, are different.
In a normal diffusion process the
slope of the variance of the soliton position is the diffusion
constant. We call this $D_{total}$ when the system is in contact with
the thermal bath since there is a contribution of both the
thermal fluctuations and the phonons. The diffusion constant due to the
noise-induced phonon bath (isolated system) is termed $D_{ph}$. 
On the other hand, our
theoretical diffusion constant, $D_{th}$, is defined as the 
linear coefficient of the Taylor expansion of $Var(x(t))$ (see Eqs.  
(\ref{appB11}) and (\ref{appB12})). Since our theory does not take into account the
contribution of phonon modes, we expect that the value
$D_{noise}=D_{total}-D_{ph}$ may be of the same order of magnitude of $D_{th}$. 
We observe in Fig. \ref{fig6} that the relative deviation of $D_{th}$
from $D_{noise}$, $((D_{noise}-D_{th})/D_{noise}$, has
the same order of magnitude of $D_{th}$ which is not 
surprising since the soliton shape is strongly
masked and distorted by the noise (see App. \ref{profiles}). In this
respect we remark  that from our results in Figs. \ref{fig3}
and \ref{fig4} we observe that
\begin{equation}\label{deviat1}
\sqrt{Var(x(t))}\,<\,L(v(t)),
\end{equation}
where the soliton width $L(v(t))$ is defined in (\ref{restframe2}).
The relation (\ref{deviat1}) means that the stochastic
deviations of the soliton center from its mean value are
relatively small compared with the soliton width. So the diffusive dynamics
of the soliton position evolves inside the soliton core. Thus, this
dynamics is very sensitive to the fluctuations of the soliton shape.
Notice that our method of determining the soliton position depends
implicitly on the soliton shape. So in the case of low-energy
solitons, where the shape is strongly masked by the noise, the
uncertainties of the method of soliton detection are relatively large compared with
the diffusive dynamics of the soliton position. This is because the diffusive
dynamics is relative small with respect to the soliton width.
In this respect we note that higher-energy solitons
present a well defined shape, i.e. the thermal fluctuations are small
respect to the soliton amplitude, so the variance of the soliton
position, given by our detection method, indeed agrees better with our
theory. (see \ref{fig3}c and \ref{fig4}c). Finally, we stress
that in our tests the phonon contribution to the soliton diffusion
could be clearly observed only for very low-energy solitons
(Figs. \ref{fig3} and \ref{fig4}, cases (c) and (e)), for higher-energy solitons the
effect is negligible, namely $D_{ph}\simeq 0$.


\subsection{Estimate of the bath temperature in real physical systems}

To clarify the physical meaning of the obtained results we have
estimated the soliton energy and characteristic temperature of the
thermal bath for two systems where solitons are believed to play an
essential role: $\alpha$-helical proteins \cite{Perez} and crystal
inertial gases \cite{ballistic}. In both cases the Lennard-Jones (LJ) interaction
potential,
\begin{equation}\label{LJ}
\phi_{LJ}=4 E_0 \left(\left(
\frac{a}{r}\right)^{12}-\left(\frac{a}{r}\right)^{6}
\right),
\end{equation}
is used. One can estimate the  potential parameters $G$ and
$A$ of the Hamiltonian (\ref{hamiltonian}) 
by using Taylor expansion of (\ref{LJ}) around the minimum
$r=(1-2^{1/6})a$. So, taking into account only the coefficients of the
harmonic and cubic terms of this expansion, one can determine the  potential parameters,
namely 
\begin{eqnarray}\label{potparamet}
A=-\frac{21}{2^{7/6} a}\quad \textrm{and}\quad G=\frac{36\,2^{2/3}\,E_0}{a^2}.
\end{eqnarray}
By using the one-soliton solution of the Bq equation (\ref{motion1})
\cite{Pnevmatikos}  and performing an integration instead of a summation
in Eq. (\ref{hamiltonian}),
the  initial soliton energy $H(0)$ reads
\begin{equation}
H(0)=\frac{16\sqrt{3}}{245}((v/c)^2-1)^{3/2}(1+9(v/c)^2)E_0.
\end{equation}
Here $v$ and $c$ are the soliton and sound velocities, respectively.
$E_0=0.22\,eV$ for $\alpha$-helix \cite{Perez} or $E_0=1.0\times
10^{-2}\,eV$ for argon \cite{argon}. So, for example, in the cases of
$v/c$=1.003 or 1.03 we get
\begin{center}
\begin{tabular}{|c|c|c|}
\hline
 {} & $\alpha$-helix & argon\\
$v/c$ & $H(0)/k_B$ [K]  &  $H(0)/k_B$ [K]\\
\hline
1.003 & 1.4 & 0.06 \\
\hline
1.03 & 45.7 & 2.15\\
\hline
\end{tabular}
\end{center}
Here $k_B$ is the Boltzmann constant. Then, the value of the
temperature of the thermal bath can be obtained by multiplying the
values of $H(0)/k_B$ by the reduced temperature. For instance, in the
$\alpha$-helix case the reduced temperature $\overline{T}=0.061419$ (see
caption Fig. \ref{fig4}) corresponds to  
$0.1\,K$ of the thermal bath if $\,v(0)=1.003$ or $3.8\,K$ if $\,v(0)=1.03$.

\section{Summary and conclusions}

We have studied the non-equilibrium diffusion dynamics of lattice
solitons on a classical chain of atoms under thermal fluctuations,
namely soliton dynamics when the system energy is close to its
stationary value. The interaction potential between the atoms is harmonic
plus a cubic anharmonicity. The chain is coupled to a thermal bath at
a given  temperature. For that reason we have included dissipation and
noise in the discrete equations of motion of the chain. Here, it is assumed
that the energy dissipation is provided by the irreversible processes arising 
from the finite velocity of the relative displacements between particles in the
chain. Thus the dissipative term  takes the form of a hydrodynamical
damping which is extensively used in elasticity theory.
The noise term
which fulfills the fluctuation-dissipation theorem then becomes a
discrete gradient of white noise delta-correlated in  space and time.  In the
continuum approach our original discrete 
set of equations leads to a form of noisy KdV-Burgers
equation. At this point we have used a collective coordinate
approach to study the diffusion dynamics of
both position and velocity of the soliton.
The soliton position and the inverse soliton width 
have been found to be good collective coordinates to describe the
soliton diffusion. We have derived two stochastic ordinary
differential equations with multiplicative noise which have been
solved analytically using stochastic perturbation analysis. 

For low-energy solitons, whose velocities are close to the sound velocity,
our molecular dynamics simulation has confirmed our analytical
predictions. Namely, normal diffusion of lattice solitons governs
short times, while superdiffusive behavior is present for long
times. The time range of the normal diffusion depends on the initial
velocity of the soliton: it is large for velocities close to the sound
velocity and short for high velocities. The collective
coordinate approach does not take into account the noise-induced
phonon bath, however we have shown that this does not play an
important role except when the 
reduced temperature (temperature in units of the initial
soliton energy $H(0)$) of the thermal bath is high. In that regime the
soliton diffusion is normal. In this case, for a given temperature,  we
have estimated in our simulations the value of the diffusion constant due to the
noise-induced phonon bath when the system energy is
close to its
stationary value. We have subtracted this value from the full value of
the diffusion constant which is not only due to the induced phonons
but also due to the noise. The order of magnitude of the resultant
value of this subtraction is predicted by the collective coordinate approach.

Since we do not observe in our numerical results any dependence on  the size of
the system, we may expect similar results for very large systems,
i.e. for $N >> 1500$.

We provided an example by using an approximation of the Lennard-Jones
potential to determine the temperature of the thermal bath 
in the cases of $\alpha$-helical proteins  and crystal
inertial gases.

Finally, our results above point out the robustness of lattice solitons. In
fact, they can exist even for higher values of  temperature and damping
constant than those explicitly considered
in the present article. On the other hand, for lower values of the
temperature the variances of the soliton position and velocity
turn out to be very small because they scale with the temperature. So
it is very difficult to observe them.

{\bf Acknowledgements}

We acknowledge support from DLR grant Nr.: UKR-002-99.
Yu. Gaididei is grateful for the hospitality of the University of Bayreuth 
where this work was performed. E. Ar\'evalo acknowledges support from
Eur. Grad. School ``Non-Equilibrium Phenomena and Phase Transitions in
Complex Systems''. A. R. Bishop is a Humboldt Awardee at the University of
Bayreuth.

\appendix

\section{Derivation of the noise term \label{appnoise}}
The goal of this Appendix is to find the form of the noise force
which would satisfy the fluctuation-dissipation theorem.
The associated set of Langevin equations of the classical chain of
atoms under thermal fluctuations are
\begin{eqnarray}\label{ap1}
\frac{dP_n}{dt}&=&T_n
+F_n^{Noise}+F_n^{Damping},\\
\label{ap1b}
\frac{dY_n}{dt}&=&\frac{P_n}{M},
\end{eqnarray}
where 
\begin{eqnarray}\label{ap5}
T_n&=&-\frac{\partial H}{\partial Y_n}=-\frac{\partial U}{\partial Y_n},\nonumber\\
F_n^{Damping}&=&M\nu\left(\frac{dY_{n+1}}{dt}-2\frac{dY_n}{dt}+\frac{dY_{n-1}}{dt}\right)
\end{eqnarray}
and  $F^{Noise}(t)$, which satisfies the fluctuation-dissipation
theorem, is to determined. $P_n$ is the momentum,
$Y_n$ denotes longitudinal displacement from its equilibrium position, of
$n-th$ particle with mass $M$ and velocity $dY_n/dt$.
$H$ is the Hamiltonian
\begin{equation}\label{ap2}
H=K+U,
\end{equation}
where
\begin{eqnarray}
K=\sum_n \frac{P_n^2}{2\,M}
\quad , \quad 
U=\sum_n V\left[Y_{n+1}-Y_{n}\right]
\end{eqnarray} 
and $V$ is an arbitrary potential which depends on the relative
displacements $Y_{n+1}-Y_{n}$. The discrete Fourier transform of
Eqs. (\ref{ap1}) and (\ref{ap1b}) read
\begin{eqnarray}\label{ap12}
\frac{d\tilde{P}_k}{dt}&=&\tilde{T}_k-\nu\tilde{\gamma }_k\tilde{P}_k
+\tilde{F}_k^{Noise}(t)\nonumber\\
\frac{d\tilde{Y}_k}{dt}&=&\frac{\tilde{P}_k}{M},
\end{eqnarray}
where 
\begin{equation}
\tilde{\gamma }_k=2\left(1-\cos( k )\right).
\end{equation}

We define
\begin{equation}\label{ap16}
\tilde{F}_k^{Noise}(t)=\sqrt{D(k)}\tilde{\xi}_k(t),
\end{equation}
where 
$\tilde{\xi}_k(t)$ is delta-correlated white noise,
\begin{equation}\label{ap15}
\left\langle
\tilde{\xi}_k(t)\tilde{\xi}_{k^{\prime}}(t^{\prime})\right\rangle 
=D(k)\delta(t-t^{\prime})\delta_{k,-k^{\prime}},
\end{equation}
and $D(k)$ is unknown.

The associated  Fokker-Planck equation of Eqs.(\ref{ap12}) in the
Stratonovich sense takes the form
\begin{eqnarray}\label{ap18}
\partial_t \rho &=&\sum_k \bigg(-\partial_{\tilde{P}_k}(T_k\rho_k)
-\frac{\tilde{P}_{-k}}{M}\partial_{\tilde{Y}_{-k}}\rho_k+\nonumber\\
&&\nu \tilde{\gamma}_k\partial_{\tilde{P}_k}\left(\tilde{P}_k \rho_k + 
\frac{D(k)}{2\nu\tilde{\gamma}_k}
\partial_{\tilde{P}_{-k}}\rho_k \right)\bigg),
\end{eqnarray}
where
\begin{equation}\label{ap18a}
\rho=\left\langle \prod_k
\delta(\tilde{P}_k-\tilde{P}_k(t))\delta (\tilde{Y}_{k}-\tilde{Y}_{k}(t))\right\rangle .
\end{equation}
In order to determine $D(k)$ we have demanded the stationary solution of 
Eq.(\ref{ap18}) to be the Boltzmann distribution, namely
\begin{equation}\label{Boltzmann1}
\rho={\cal N} \exp\bigg(-{\frac{H}{k_BT}}\bigg),
\end{equation}
where $H$ is defined in Eq. (\ref{ap2}) and ${\cal N}$ is the
normalization constant. Substituting Eq.(\ref{Boltzmann1}) into Eq.(\ref{ap18}),
it is straightforward to see that 
\begin{equation}\label{ap19}
D(k)=2\nu\tilde{\gamma}_k\,M\,k_B\,T.
\end{equation}
Therefore, from Eq. (\ref{ap16}) together with (\ref{ap19}), it is
easy to show that in  position space
\begin{eqnarray}\label{ap21}
&&\left\langle
F_n^{Noise}(t)F_{n^{\prime}}^{Noise}(t^{\prime})\right\rangle 
=\nonumber\\
&&-2\nu \,M\,k_B\,T \delta(t-t^{\prime})(\delta_{n+1,n^{\prime}}
-2\delta_{n,n^{\prime}}+\delta_{n-1,n^{\prime}}).\quad\quad\quad
\end{eqnarray}
Finally, the relation (\ref{ap21}) can be satisfied by the definition 
\begin{equation}
F_n^{Noise}(t)=\sqrt{2\nu \,M\,k_B\,T}(\xi_{n+1}(t)-\xi_n(t)),
\end{equation}
where
\begin{equation}
\left\langle \xi_n(t)\xi_{n^{\prime}}(t^{\prime})\right\rangle 
=\delta(t-t^{\prime})\delta_{n,n^{\prime}}.
\end{equation}

\section{Continuum limit}

In order to reduce Eq. (\ref{motion0}) to a form of noisy and damped
KdV equation we have performed two steps. First, we have employed the
continuum approach \cite{Pnevmatikos,Remoissenet} in order
to obtain a form of noisy and damped Bq equation, and then we have
used the reductive perturbation technique \cite{Taniuti,Remoissenet} in
order to obtain the noisy and damped KdV equation.

\subsection{Noisy and damped Bq equation}

Here we have used the procedure of Pnevmatikos \cite{Pnevmatikos}, who expanded
$Y_{n\pm 1}(t)$ and $Y_{n\pm 2}(t)$ in a Taylor series around
$y(x,t)$, with $x=na$, where the equilibrium atomic spacing $a$ is
regarded as an expansion parameter. Then, collecting powers of $a$,
Eq. (\ref{motion0}) together with 
Eqs. (\ref{damping1}) and (\ref{noiseespecial}) at $O(a^4)$ takes the form
\begin{equation} \label{motion1}
\partial_{t}^2 y=c^2 \partial_{x}^2 y + p  \partial_{x}y
\partial_{x}^2y +h \partial_{x}^4y
+\nu a^2\partial_{x}^2\partial_{t} y+ a \sqrt{{\it D}} \partial_{x}\xi(x,t) ,
\end{equation} 
where 
\begin{equation}
\frac{\xi_{n+1}(t)-\xi_n(t)}{a^{3/2}}\rightarrow\partial_{x}\xi(x,t),
\end{equation}
with properties
\begin{eqnarray}\label{proper1}
\left\langle \partial_{x}\xi(x,t)\right\rangle &=&0,\nonumber\\
\left\langle \partial_{x}\xi(x,t)\partial_{x^{\prime}}\xi(x^{\prime},t^{\prime})\right\rangle 
&=&\partial_{x^{\prime}}\partial_{x}\delta(x-x^{\prime})\delta(t-t^{\prime}).\quad\quad
\end{eqnarray}
The diffusion constant takes the form
\begin{equation}
{\it D}=\frac{2\nu k_BT}{\rho}.
\end{equation}
Other constants are
\begin{eqnarray}\label{constants1} 
c^2 =\frac{Ga}{\rho},
\quad && \quad
p  =  \frac{2a^2AG}{\rho},\nonumber\\
h =\frac{a^3G}{12 \rho},
\quad && \quad
\rho =  M/a. 
\end{eqnarray}

\subsection{Noisy and damped KdV equation}

We write  Eq. (\ref{motion1}) in the sound velocity frame and make
a further approximation concerning variations in time. In this case we
may use the reductive perturbation technique \cite{Taniuti}. Therefore we
rewrite  Eq. (\ref{motion1}) in the  perturbation form
\begin{eqnarray} \label{motion2}
&&\partial_{t}^2 y-c^2 \partial_{x}^2 y - p  \partial_{x}y
\partial_{x}^2y -h \partial_{x}^4y=\nonumber\\
&&\quad\quad\quad\quad\kappa\left(
\nu a^2\partial_{x}^2\partial_{t} y+ a \sqrt{{\it D}} \partial_{x}\xi(x,t)\right) ,
\end{eqnarray} 
where we have introduced a small parameter $\kappa$ for
convenience. Afterwards, we perform the following change of variables
\begin{equation}\label{galileo1}
s= \kappa\alpha (x-c t), \quad\quad\quad \tau= \kappa^3\beta t, \quad\quad\quad 
u=\gamma \partial_s y, 
\end{equation}
so that
\begin{eqnarray}\label{galileo1a}
\partial_x=\kappa\alpha\partial_s\quad \textrm{and}
\quad\partial_t=\kappa^3\beta \partial_{\tau}-\kappa\alpha c\partial_s,
\end{eqnarray}
where
\begin{equation}\label{galileo2}
\alpha=\frac{p }{\sqrt{6h}},\quad\quad
\beta=\frac{p ^3}{12c\sqrt{6h}},\quad\quad \gamma=\frac{1}{\sqrt{6h}}.
\end{equation}
We have also expressed  $u$ and $\xi$ in a perturbation series
\begin{eqnarray}\label{uexpan}
u=\kappa u_1 + \kappa^2 u_2 + \cdot\cdot\cdot\\
\label{xiexpan}
\xi=\kappa \xi_1 + \kappa^2 \xi_2 + \cdot\cdot\cdot
\end{eqnarray}
Here the parameter $\kappa$ indicates the magnitude of the rate of
change, the coefficients $\kappa$ and $\kappa^3$ in
(\ref{galileo1}) are chosen in order to balance the nonlinear term,
and the dispersive term and the 
time derivative are of the same order in $\kappa$. The small noise
expansion (\ref{xiexpan}) is defined such that the lowest order
terms of noise and damping are of the same order.
Substituting Eqs. (\ref{galileo1a}), (\ref{uexpan}) and  (\ref{xiexpan})
into (\ref{motion2}), and keeping the 
lowest order terms, namely $O(\kappa^5)$, we find that
\begin{equation}
\partial_{\tau} u+6u\partial_s u+\partial_s^3 u
=\nu_1 \partial_{ss}u-\sqrt{D_1}\partial_s \xi(s,\tau),
\end{equation}
where we have set $u=u_1$ and $\xi=\xi_1 $.
\begin{eqnarray}\label{kdv01}
\nu_1=\frac{\sqrt{6}\nu a^2 c}{\sqrt{h}p },\quad
D_1={\it D}\alpha\,\beta\left(\frac{6\,a}{p ^3}\right)^2.
\end{eqnarray}

\section{Perturbation analysis}
\begin{figure*}
\includegraphics{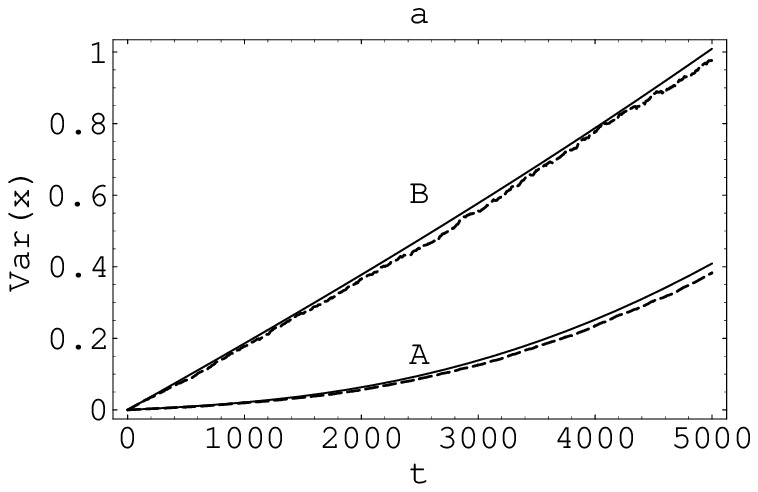}
\includegraphics{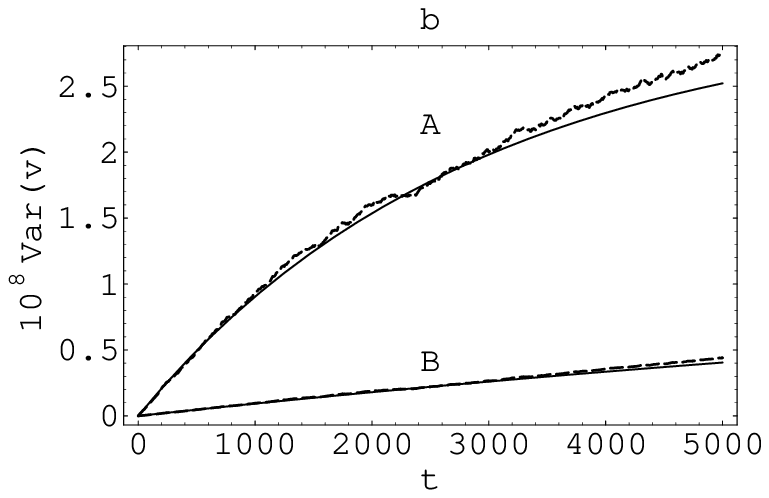}
\includegraphics{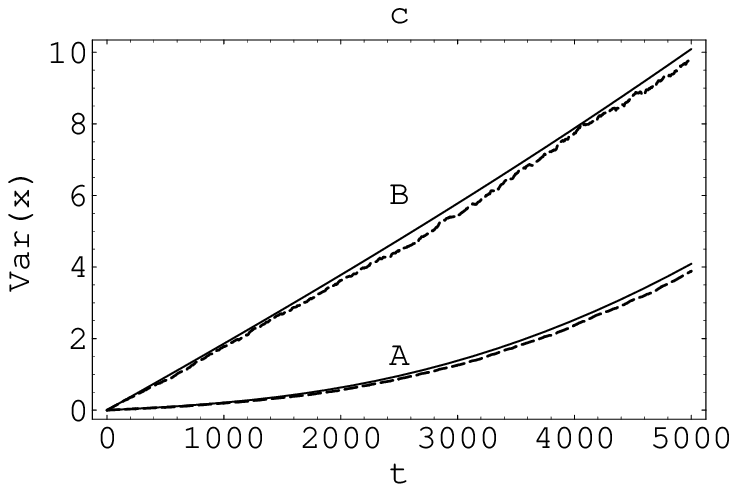}
\includegraphics{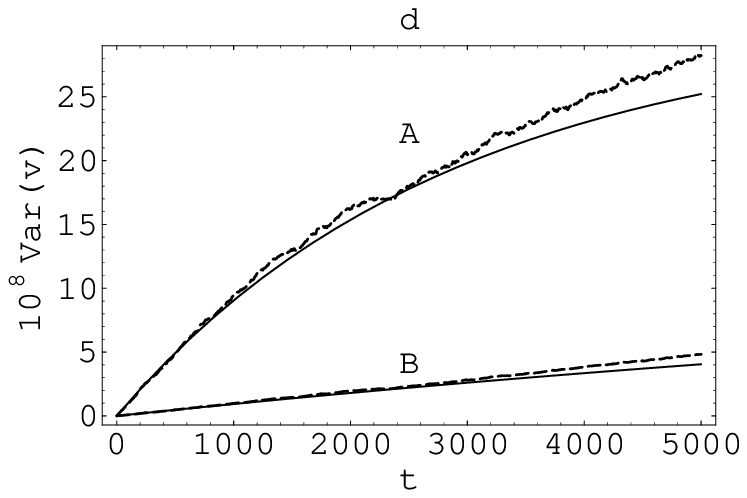}
\caption{$Var(x)$ (Eq. (\ref{appB12})) and $Var(v)$ (Eq.(\ref{vel2}))
vs. time compared with results from  a numerical solution of
Eqs. (\ref{collective6}) and (\ref{collective7}). 
In  panels (a) and (b) $T=5\times 10^{-6}$, and in  panels (c) and (d)
$T=5\times 10^{-5}$. A and B in all cases correspond to
$v(0)=1.005$ and $v(0)=1.001$, respectively. $\nu\,=\,0.003$. Solid
line: analytical prediction, dashed line: numerical solution.}
\label{figap1}
\end{figure*}

In this Appendix we develop a perturbation approach to the equations
(see Eqs (\ref{collective8}) and (\ref{collective9})) 
\begin{eqnarray}\label{appB1}
dS(\tau)&=&4\eta^2(\tau)d\tau+\epsilon
\frac{5\sqrt{3}}{4\sqrt{7}}\sqrt{\frac{D_1}{\eta^3(\tau)}}\,dW_1(\tau)\\
\label{appB1a}
d\eta(\tau)&=&-\frac{30\nu_1}{30+\pi^2}\eta^3(\tau)d\tau+\epsilon\bigg(
\frac{225(231+8\pi^2 )D_1}{112(30+\pi^2)^2}d\tau+\nonumber\\
&&
\frac{15\sqrt{21+\pi^2}}{2\sqrt{7}(30+\pi^2)}\sqrt{D_1\eta(\tau)}\,dW_2(\tau)\bigg).
\end{eqnarray}
We interpret Eqs. (\ref{appB1}) and (\ref{appB1a}) in the Ito sense
where the Wiener process $dW_i(\tau)=\xi_i(\tau)d\tau$ with
$\langle\xi_i(\tau) \xi_j(\tau^{\prime})\rangle=\delta_{ij}\delta(\tau-\tau^{\prime})$.
We seek an asymptotic
solution of the form
\begin{eqnarray}\label{appB2}
S(\tau)=s_0(\tau)+\epsilon\,s_1(\tau)+\cdot\cdot\cdot\nonumber\\
\eta(\tau)=\eta_0(\tau)+\epsilon\,\eta_1(\tau)+\cdot\cdot\cdot.
\end{eqnarray}
Inserting Eqs. (\ref{appB2}) into Eqs. (\ref{appB1}) and
(\ref{appB1a}) and collecting powers of $\epsilon$ we get\\
$\epsilon^0$:
\begin{eqnarray}\label{appB3}
ds_0(\tau)&=&4\eta_0^2(\tau)d\tau\\
\label{appB4}
d\eta_0(\tau)&=&-\frac{30\nu_1}{30+\pi^2}\eta_0^3d\tau
\end{eqnarray}
$\epsilon^1$:
\begin{eqnarray}\label{appB5}
ds_1(\tau)&=&8\,\eta_0(\tau)\,\eta_1(\tau)d\tau+
\frac{5\sqrt{3}}{4\sqrt{7}}\sqrt{\frac{D_1}{\eta_0^3(\tau)}}\,dW_1(\tau)\quad\quad\quad\\
\label{appB6}
d\eta_1(\tau)&=&-\frac{90\nu_1}{30+\pi^2}\eta_0^2(\tau)\eta_1(\tau)d\tau+\nonumber\\
&&
\frac{15\sqrt{21+\pi^2}}{2\sqrt{7}(30+\pi^2)}\sqrt{D_1\eta_0(\tau)}\,dW_2(\tau).
\end{eqnarray}
Solving Eqs. (\ref{appB3}) and (\ref{appB4}) we obtain
\begin{eqnarray}\label{appB7}
s_0(\tau)&=&4\frac{\eta_0^2(0)}{\lambda}\log{(1+\lambda \tau)}\\
\label{appB8}
\eta_0(\tau)&=&\frac{\eta_0(0)}{\sqrt{1+\lambda \tau}}
\end{eqnarray}
with
\begin{eqnarray}\label{appB81}
\lambda=\frac{60\nu_1\eta_0^2(0)}{30+\pi^2}.
\end{eqnarray}
Inserting Eq. (\ref{appB8}) in  (\ref{appB6}) and solving, with the
initial condition $\eta_1(0)=0$, we find
\begin{eqnarray}\label{appB9}
&&\eta_1(\tau)=\frac{45\,D_1(231+8\pi^2)((1+\lambda\,\tau)^{5/2}-1)}
{56(30+\pi^2)^2\lambda(1+\lambda\,\tau)^{3/2}}+\nonumber\\
& &\frac{15\sqrt{21+\pi^2}}{2\sqrt{7}(30+\pi^2)}\sqrt{D_1\eta_0(0)}
\int\limits_{0}^{\tau}(1+\lambda \tau^{\prime})^{5/4}\,dW_2(\tau^{\prime}).\quad\quad\quad\nonumber\\
\end{eqnarray}
Then inserting Eqs. (\ref{appB9}) and  (\ref{appB8}) in Eq
(\ref{appB5}) and integrating once we get
\begin{eqnarray}\label{appB10}
&&s_1(\tau)=\frac{15D_1(231+8\pi^2)\,\eta_0(0)}
{7(30+\pi^2)^2\lambda^2(1+\lambda\tau)}\,\bigg(2(1+\lambda\tau)^{5/2}-\quad\quad\quad\nonumber\\
&&\quad\quad 5\lambda\tau-2\bigg)+
\frac{5\sqrt{3}\sqrt{D_1}}{4\sqrt{7}\eta_0^{3/2}(0)}
\int\limits_{0}^{\tau} 
(1+\lambda \tau^{\prime})^{3/4}
\,dW_1(\tau^{\prime})+\nonumber\\
&&\quad\quad\frac{60\sqrt{21+\pi^2}}{\sqrt{7}(30+\pi^2)}\sqrt{D_1}\,\eta_0^{3/2}(0)\times\nonumber\\
&&\quad\quad\int\limits_{0}^{\tau} d\tau^{\prime}
\frac{1}{\sqrt{1+\lambda \tau^{\prime}}}
\int\limits_{0}^{\tau^{\prime}}
(1+\lambda
\tau^{\prime\prime})^{5/4}\,dW_2(\tau^{\prime\prime}).
\end{eqnarray}
To this order we have $Var(S(\tau))=\epsilon^2\,Var(s_1(\tau))$ and
thus we finally obtain
\begin{eqnarray}\label{appB11}
Var(S(\tau)) &=&\epsilon^2\bigg(
\frac{75\,D_1}{112\eta^3_0(0)}\tau +\frac{225\,D_1\lambda}{448\eta^3_0(0)}\tau^2
+ O(\tau^3)\bigg).\nonumber\\
\end{eqnarray}
The full expressions of $Var(S)$ to this order of perturbation is
given by Eqs. (\ref{pertur2}).
Notice that the variance in the rest frame reads
\begin{equation}\label{appB12}
Var(x(t))=\frac{1}{\alpha^2}Var(S(\beta\,t)).
\end{equation}
Concerning soliton velocity, up to first order
perturbation it reads
\begin{equation}\label{vel1}
4\,\eta^2(\tau)=4\eta_0^2(\tau)+8\epsilon\,\eta_0(\tau)\,\eta_1(\tau).
\end{equation}
Then, by substituting Eqs. (\ref{appB8}) and (\ref{appB9}) in
Eq. (\ref{vel1}), it is straightforward to see that
\begin{eqnarray}\label{appB13}
&&Var(4\,\eta^2(\tau)) = 64\,\epsilon^2\,\langle\eta_0(\tau)\rangle^2 Var(\eta_1^2(\tau))\nonumber\\ 
&&\quad\quad\quad=
\epsilon^2\,\bigg(\frac{3600(21+\pi^2)\,D_1}{7(30+\pi^2)^2}\eta^3_0(0)\tau-\nonumber\\ 
&&\quad\quad\quad
\frac{9900(21+\pi^2)\,D_1\lambda}{7(30+\pi^2)^2}\eta^3_0(0)\tau^2
+ O(\tau^3)\bigg).\quad\quad\quad
\end{eqnarray}
The full expressions of $Var(4\,\eta^2)$ to  this order of perturbation
is given by Eqs. (\ref{pertur2}). In the rest frame this variance reads
\begin{equation}\label{vel2}
Var(v(t))=\left(\frac{p^2}{12\,c}\right)^2\, Var(4\,\eta^2(\beta\,t)).
\end{equation}
In Fig. \ref{figap1} we show some examples of $Var(x)$ and $Var(v)$
compared with results from a numerical solution of
Eqs. (\ref{collective6}) and (\ref{collective7}) for which we have used the 
Heun method \cite{Heun}. The variances have been obtained by averaging
over 1000 realizations.

\section{Thermalization process \label{thermalization}}

\begin{figure}
\includegraphics{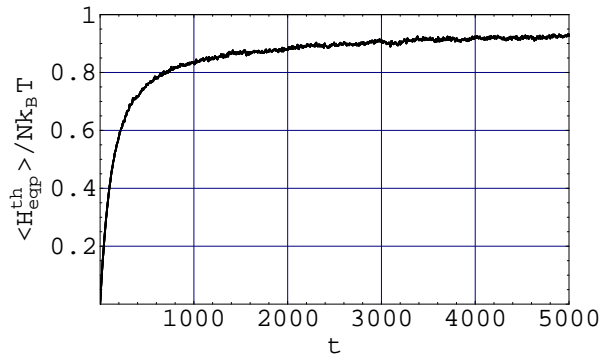}
\caption{$\langle H^{th}_{epq}\rangle /Nk_BT$ vs. t. $\nu=0.003$.}
\label{figequip}
\end{figure}

\begin{figure*}
\includegraphics{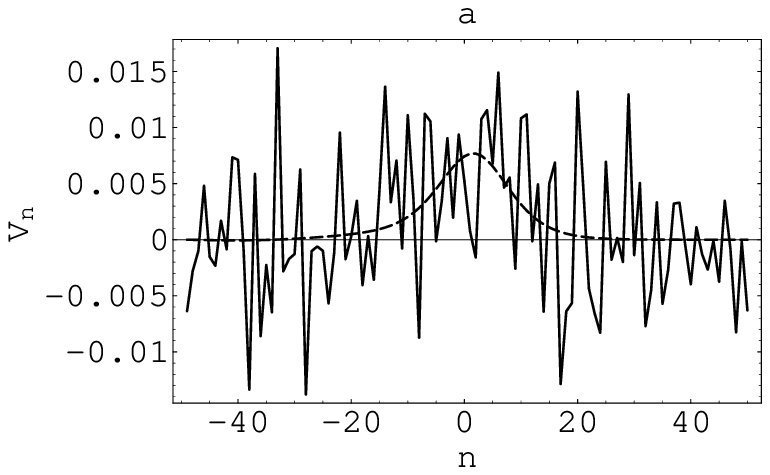}
\includegraphics{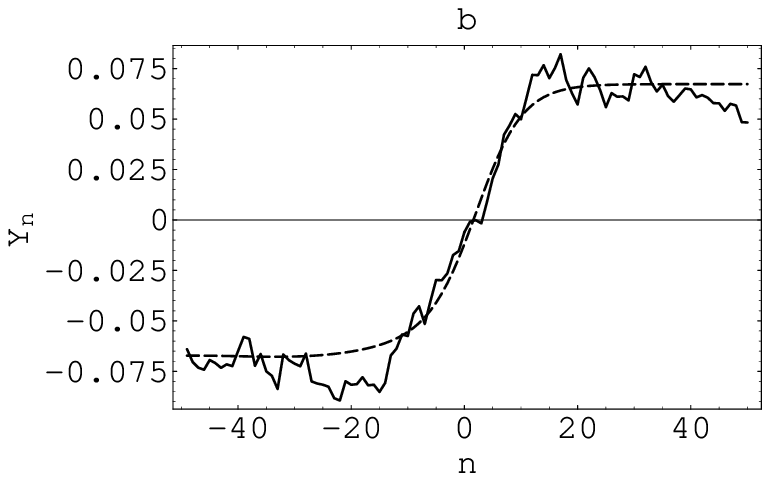}
\caption{Lattice soliton profiles: pulse shape (a), kink shape (b). 
Solid line: noisy shape, Dashed line: shape in the presence of damping
only.
$T=5\times 10^{-5}$, v(0)=1.003, $\nu=0.003$ and $t=5000$}
\label{fig7}
\end{figure*}

From the generalized equipartition theorem \cite{Huang} we have that 
\begin{equation}\label{equip0}
\left\langle \sum_{n=1}^{N}
Y_n(t)\frac{\partial H(t)}{\partial Y_n(t)} \right\rangle=N\,k_B\,T
\end{equation}
when the system is in thermal equilibrium with an external bath at
temperature $T$. The relation (\ref{equip0}) is strictly satisfied
in the harmonic limit of the Hamiltonian (\ref{hamiltonian}), namely
$A=0$. For finite values of $A$ 
the relation (\ref{equip0}) is a rather good approximation to evaluate the
temperature of our system when the relative displacements are sufficiently
small, namely
\begin{equation} \label{condit1}
Y_{n+1}(t)-Y_n(t)<<\frac{3}{2A}.
\end{equation}
Notice that $A=1$ in our simulations.
The condition (\ref{condit1}) can be obtained by
comparing the harmonic term with the cubic anharmonicity in
(\ref{hamiltonian}) and  it is always satisfied in
the present work.

So, in order to  examine the thermalization process,
one can define the ensemble average
\begin{equation}\label{equip1}
\langle H_{eqp}(t)\rangle=\left\langle \sum_{n=1}^{N}
Y_n(t)\frac{\partial H(t)}{\partial Y_n(t) } \right\rangle.
\end{equation}
For finite times  $\langle H_{eqp}(t)\rangle$ possesses two
contributions, one due to the coupling to the thermal bath, $\langle
H_{eqp}^{th}(t)\rangle$,  and the
other one due to the soliton, $\langle H_{eqp}^{sol}(t)\rangle$. So 
\begin{equation} \label{sumequip1}
\langle H_{eqp}(t)\rangle=\langle H_{eqp}^{th}(t)\rangle
+\langle H_{eqp}^{sol}(t)\rangle ,
\end{equation}
Notice that for a very large system ($N>>1500$) the contribution of the
soliton energy, $\langle H_{eqp}^{sol}(t)\rangle$, can be
neglected. However, in our system of 1500 sites 
this contribution is appreciable. In fact, the time evolution of
$\langle H_{eqp}(t)\rangle$ for different initial soliton energies
presents different values due to the soliton
contribution. Here,  
we remark that in thermal equilibrium, i.e. $t\rightarrow \infty$,
these differences vanish. In Eq. (\ref{sumequip1})
$\langle H_{eqp}^{th}(t)\rangle$ gives us
information about the time evolution of the temperature of the
system in terms of the temperature of the thermal bath. The soliton
contribution $\langle H_{eqp}^{sol}(t)\rangle$ can be evaluated
numerically using Eq. (\ref{equip1}) when the soliton propagates in
the presence of the damping but without noise. Then one can perform
the numerical subtraction $\langle H_{eqp}(t)\rangle- \langle
H_{eqp}^{sol}(t)\rangle$  to obtain $\langle H_{eqp}^{th}(t)\rangle$,
which is shown in a  normalized form in Fig. \ref{figequip}. We
remark that we observe the same result here
in systems with the double number of sites, namely 3000. The
normalized thermalization process depends only on 
the damping constant which has the same value $\nu=0.003$ in all our
simulations. Notice that for
times $t\, {}^{<}_{\sim}\,3/\nu=1000$ there is a fast relaxation process,
but for larger times, $t\, {}^{>}_{\sim}\,3/\nu$,  the energy system approaches
very slowly its stationary value. The thermal equilibrium of the system
with the external bath corresponds to the case
$\langle H_{eqp}^{th}(t)\rangle/k_BT=1$.\\

\section{Profiles \label{profiles}}

In Fig. \ref{fig7} we compare snapshots of the system at $t=5000$
with and without noise and in the presence of
damping. Figs.  \ref{fig7}a and  \ref{fig7}b correspond to both the kink
shape (absolute displacements)  and the pulse shape (relative displacements),
respectively. Both shapes, with and without
noise, in Fig. \ref{fig7}b were reconstructed from the shapes in
Fig. \ref{fig7}a, respectively, by using the algorithm (\ref{algorithm}).


\begin{thebibliography}{99}

\bibitem{Lomdahl} P.S. Lomdahl, W. C. Kerr, Phys. Rev. Lett., {\bf
55}, 1235 (1985).

\bibitem{Lawrence} A. F. Lawrence, J. C. McDaniel, D. B. Chang,
B. M. Pierce, R. R. Birge, Phys. Rev. A {\bf 33}, 1188 (1986).

\bibitem{Toda}  M. Toda,  {\it Theory of Nonlinear Lattices}
(Spring-Verlag, 1981).

\bibitem{Pnevmatikos} {\it Singularities $\&$ Dynamical Systems}, edited by S. N. 
Pnevmatikos (North-Holland, 1985).

\bibitem{Davydov} A. S. Davydov, 
{\it solitons in Molecular Systems} (Reidel, 1985).

\bibitem{Collins} M. A. Collins, Chem. Phys. Lett. {\bf 77},342 (1981)

\bibitem{Hochstrasser1} D. Hochstrasser, F. G. Mertens and
H. B\"uttner, Phys. Rev. A {\bf 40}, 2602 (1989).

\bibitem{Yomosa} S. Yomosa, Phys. Rev. A {\bf 32}, 1752 (1985).
 
\bibitem{Perez} P. Perez and N. Theodorakopoulos, Phys. Lett. A {\bf
117}, 405 (1986).

\bibitem{Perez1} P. Perez and N. Theodorakopoulos, Phys. Lett. A {\bf
124}, 267 (1987).

\bibitem{Neuper} A. Neuper and F. G. Mertens, in {\it Nonlinear
Excitations in Biomolecules}, edited by M. Peyrard (Les Editions de
Physique, Springer, 1995).

\bibitem{Muto} V. Muto, P.S. Lomdahl and P.L. Christiansen,
Phys. Rev. A {\bf 42}, 7452 (1990).

 
\bibitem{samsonov} A. M. Samsonov, {\it Strain Solitons in Solids and
How to Construc them} (Chapman \& Hall/CRC, 2001)

\bibitem{KonotopVazquez} V. Konotop
and L. Vazquez, {\it Nonlinear Random waves} (World Scientific, 1994).

\bibitem{Wadati} M. Wadati, J. Phys. Soc. Jpn. {\bf 52}, 2642 (1983).

\bibitem{Wadati2} M. Wadati and Y. Akutsu, J. Phys. Soc. Jpn. {\bf 53}, 3342
(1984).



\bibitem{Herman} R. L. Herman, J. Phys. A {\bf 23}, 1063 (1990).

\bibitem{Iizuka} T. Iizuka, Phys. Lett A {\bf 181}, 39 (1993).

\bibitem{ScalerandiRomano} M. Scalerandi, A. Romano and C. A. Condat, Phys. Rev. E
{\bf 58} 4166 (1998).

\bibitem{Remoissenet} M. Remoissenet  {\it Waves Called Solitons}
(Springer,1996).

\bibitem{kampetter} T. Kamppeter, F.G. Mertens, E. Moro, A. Sanchez,
A. R. Bishop, Phys. Rev. B {\bf 59}, 11349 (1999). 


\bibitem{mathias1}   M. Meister, F. G. Mertens, J. Phys. A {\bf 33},
2195 (2000). 


\bibitem{mathias} M. Meister, F. G. Mertens, and A. S\'anchez,
Eur. Phys. B {\bf 20}, 405 (2001).

\bibitem{Landau} L. D. Landau and 
E. M. Lifshitz, {\it Theory of Elasticity} (Pergamon Press, 1986).

\bibitem{Schmittmann} B. Schmittmann and R. K. P. Zia, {\it
Statistical Mechanics of Driven Diffusive Systems} (Academic Press, 1995).

\bibitem{nozaki} K. Nozaki, J. Phys. Soc. Jpn. {\bf 56}, 3052 (1987).

\bibitem{Orlowski} A. Orlowski, Phys. Rev. E {\bf 49}, 2465 (1994).

\bibitem{arevalo} E. Ar\'evalo, Yu. Gaididei and F. G. Mertens,
Eur. Phys. J. B {\bf 27}, 63 (2002). 

\bibitem{grim} R. Grimshaw and H. Mitsudera, Stud. Appl. Math.
{\bf 90}, 75 (1993).

\bibitem{niurkaPRE} N. R. Quintero, A. S\'anchez, F. G. Mertens,
Phys. Rev. E {\bf 62}, 5695 (2000). 

\bibitem{Mertens1} F. G. Mertens, H. J. Schnitzer, and A. R. Bishop,
Phys. Rev. B {\bf 56}, 2510 (1997).



\bibitem{Mann} E. Mann, J. Math. Phys. {\bf 38}, 3772 (1997).


\bibitem{Gardiner} C. W. Gardiner, {\it Stochastic Methods}
(Springer-Verlag, 1990). 

\bibitem{Abdullaev} F. Kh. Abdullaev, Physics Reports {\bf 179}, 1
(1989).

\bibitem{Heun} Peter E. Kloeden, Eckhard Platen, {\it Numerical solution of
Stochastic Differential Equations} (Springer-Verlag, 1992).


\bibitem{niurka1} N. R. Quintero, A, S\'anchez, F. G. Mertens,
Eur. Phys. B {\bf 16}, 361 (2000).

\bibitem{ballistic} A. Cenian and H. Gabriel, J. Phys.: Condens. Matter
{\bf 13} 1 (2001).

\bibitem{argon} K. Laasonen, S. Wonczak, R. Strey and A. Laaksonen
J. Chem. Phys. {\bf 113} 9741 (2000).

\bibitem{Taniuti} Taniuti T. and Wei C., Phys. Soc. Jap. {\bf 21}, 209
(1968).


 



\bibitem{Huang} K. Huang, {\it Statistical Mechanics}, 2nd ed. (John
Wiley and Sons, 1987).

\end{thebibliography}
\end{document}